\pdfoutput=1
\documentclass{JINST}
\let\ifpdf\undefined
\usepackage{amsmath}
\usepackage{textcomp} 

\usepackage{ifpdf}

\DeclareGraphicsExtensions{.pdf,.png,.jpg} 

\title{NIKEL: Electronics and data acquisition for kilopixels kinetic inductance camera}
\author{O.~Bourrion$^a$\thanks{Corresponding author.}, 
C.~Vescovi$^a$,
J.L.~Bouly$^a$,
A.~Benoit$^b$,
M.~Calvo$^b$,
L.~Gallin-Martel$^a$,
J.F.~Macias-Perez$^a$,
A.~Monfardini$^b$
.\\
\llap{$^a$}Laboratoire de Physique Subatomique et de Cosmologie,\\ 
Universit\'e Joseph Fourier Grenoble 1,\\
  CNRS/IN2P3, Institut Polytechnique de Grenoble,\\
  53, rue des Martyrs, Grenoble, France\\
\llap{$^b$}Institut Néel, CNRS/UJF, \\
  25 rue des Martyrs, Grenoble, France \\
}  

\abstract{
A prototype of digital frequency multiplexing electronics allowing the real time monitoring of microwave kinetic inductance detector (MKIDs) arrays for mm-wave astronomy has been developed. 
Thanks to the frequency multiplexing, it can monitor simultaneously 400 pixels over a 500\,MHz bandwidth and requires only two coaxial cables for instrumenting such a large array. 
The chosen solution and the performances achieved are presented in this paper.
} 


\begin{document}

\section{Introduction}
Microwave kinetic inductance detectors (MKIDs) have proven to be a solid working alternative to traditional bolometers for millimeter and sub-millimeter astronomy \cite{Monfardini,Monfardini2011,Baselmans,Schlaerth}. 
MKIDs are composed of high-quality superconducting resonant circuits electromagnetically coupled to a transmission line. 
They are designed to resonate in the microwave domain \cite{Day,Mazin2,DoyleThesis}. 
For astronomical applications, the resonances typically lie between 1 to 10\,GHz and have loaded quality factors around $\rm Q_{L}=10^5$, corresponding to a typical bandwidth of $\rm \Delta f = f/Q_{L} \sim  10-100$\,kHz. Provided that the MKID resonant frequencies can be easily adjusted by layout design, it is possible to couple a large number of MKIDs with different resonance frequencies to a single transmission line \cite{Swenson}.  Indeed, a large number of MKIDs can naturally be read out by a frequency-based multiplexing system with no loss of performance \cite{Bourrion2011}.
In practice, the average frequency spacing between resonators is between 1 and 2\,MHz \cite{Monfardini}.
Thus, in order to ensure the largest sky coverage and overall signal to noise per unit of time with a reduce number of cables (few) feedthrough to the cryostat, the analog bandwidth and the number of detectors (resonators) managed by the electronics must be maximized. 
At this respect, we present here a building block for the NIKA camera \cite{Monfardini,Monfardini2011} that is able to monitor simultaneously 400 pixels over a 500\,MHz bandwidth. 

\section{Instrumentation methodology}
\label{InstruMetho}
The instrumentation setup used for NIKA and its associated electronics is extensively described in \cite{Bourrion2011}. 
In summary, the excitation frequency comb is generated at baseband in the electronics using coordinate rotation digital computer (CORDIC), up-converted with an IQ mixer to the 1 to 10\,GHz frequency and injected in the resonator line. 
The returning and thus modified frequency comb is down-converted and analyzed by channelized Digital Down Converters (DDC) to determine each tone amplitude and phase.
Aside from good signal to noise ratio (SNR) on the whole chain, the first limitation on the number of MKIDs managed by this solution is given by the  digital to analog converter (DAC) and the analog to digital converter (ADC) bandwidths. 
The second constraint comes from the computing power limitation. For a FPGA (Field Programmable Gate Array), the computational power is determined by the available amount user logic and multiplier block times their maximum running frequency. Indeed, thanks to the inherently achievable parallelization in FPGAs, this figure is much larger compared to DSPs that have only a few Multiplier Accumulators.

\begin{figure}[th]
\begin{center}
\includegraphics[angle=-90,width=12cm]{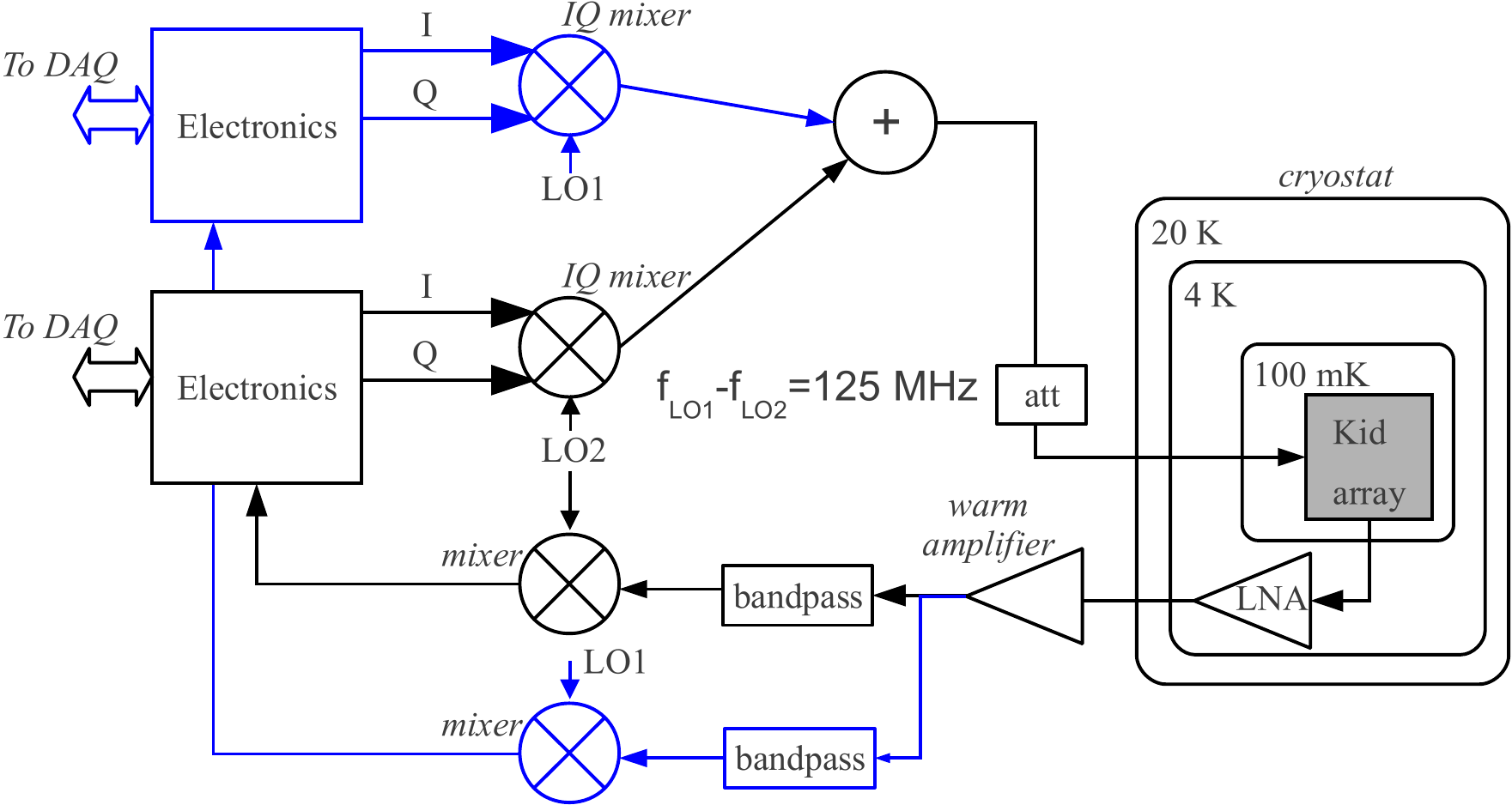}
\caption{Overview of the setup using the juxtaposition of several electronics to monitor a MKID array. Each electronics generating the two frequency combs (each tone phase shifted by 90\textdegree\  between I and Q) is followed by an IQ up-mixer. The excitation combs up-converted at high frequencies are summed and the resulting signal is fed to a programmable attenuator for power adjustment. After passing through the cryostat and the low noise boost amplifier each share of bandwidth is separated by highly selective filters before passing through the  down-mixers and returned to the corresponding electronics.}
\label{AllAnalogSolution}
\end{center}
\end{figure}

Starting from the previous version, which was able to manage a line of 128 tones over a bandwidth of 125\,MHz, three solutions are possible to increase the multiplexing factor per line. 
The first solution would be to juxtapose several of the previous electronic boards, each one managing its share of bandwidth, see figure~\ref{AllAnalogSolution}. 

Unfortunately, the analog filters required to separate each share of bandwidth before down-converting have such a stringent separation requirement to avoid crosstalk due to image frequencies that they cannot be constructed.

The second option is to use faster ADCs an DACs combined to a larger computing power (FPGA) in order to directly cover a larger bandwidth. Following this path, two concurrent approaches still remain. 
The first ``obvious'' solution is to directly generate the frequency comb at twice the desired bandwidth and to perform channelized DDC with the ADC signal. 
Unfortunately, due to the frequency limitation of state of the art FPGAs this can only be achieved by performing massive design pipelining on both sides, excitation and analysis, and therefore makes it extremely complicated. 

The third option, which we have chosen, is to use modern DACs featuring digital modulator and interpolator followed by very steep half-band filters for generating the excitation comb.
With these, the total frequency bandwidth to cover is split into smaller bands where the frequency combs can be computed at a moderate frequency, digitally up-converted and filtered to avoid unwanted spurious frequencies. Finally, each band contribution is then summed before being up-converted to the frequency band of interest by an IQ up-mixer. 
At reception side, the returning signal is down converted to baseband and is digitized by a fast ADC. Then, the digitized signal goes through a polyphase filter bank with equal bandwidth overlapping bands. This filter, has the ability to separate the total bandwidth in five smaller frequency bands and to down convert each of them to baseband. The sub frequency bands are chosen such as to match the excitation bands.
The filter outputs are fed to the corresponding channelized DDC in order to be analyzed.
The benefit of this architecture is to limit the massive pipelining to the polyphase filter part, and thus, to dramatically reduce the required amount of user logic for the frequency comb generation and the 
channelized DDC.

\section{Hardware development}
\label{HardwareDevel}
\begin{figure}[th]
\begin{center}
\includegraphics[angle=0,width=0.95\textwidth]{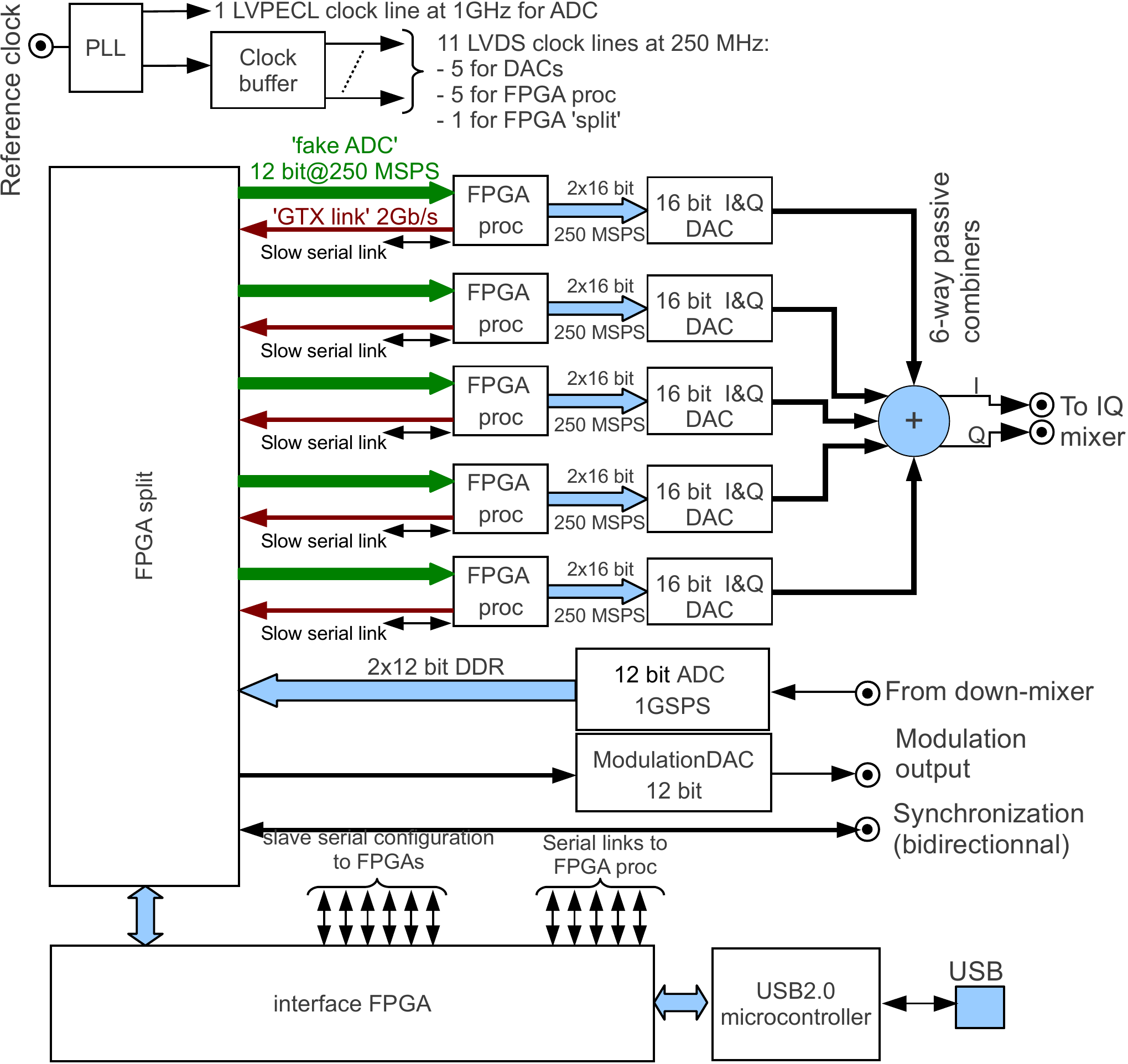}
\caption{The electronic board is composed of a central FPGA (labeled `split') which receives the  12 bit ADC (ADS5400 from Texas Instruments) output data flow at 1\,GSPS and of five processing FPGA (labeled `proc'). 
Each of the latter is driving its associated DAC with the adequate frequency comb which can feature up to 80 tones. The `proc' FPGA is connected to the `split' FPGA with two links. The first of these, labeled `fake ADC', is a 12 bit parallel LVDS link running at 250\,MSPS that is carrying the part of bandwidth corresponding to the excitation signal. 
The second link, labeled `GTX link', is periodically (at $\sim$953\,Hz) conveying the 80 DDC results over a 2\,Gb/s serial link.
An additional slow speed DAC, driven by the `split' FPGA, is implemented to be able to provide a 500\,Hz modulation signal. The communication with the hardware is ensured via a USB2 capable micro-controller and an interface FPGA that accommodates different voltage levels.}
\label{HardwareOverview}
\end{center}
\end{figure}

Following section~\ref{InstruMetho}, a dedicated hardware,  the New Iram KID ELectronics (NIKEL), able to manage 400 resonators over a bandwidth of 500\,MHz was developed. NIKEL is designed such as to manage five adjacent bands of 100\,MHz. 
This choice was driven by the chosen DAC capabilities (AD9125 from analog devices). As shown in figure~\ref{HardwareOverview}, the NIKEL electronic board is composed of a central FPGA (labeled `split') which receives the 12 bit ADC (ADS5400 from Texas Instruments) output data flow at 1\,GSPS and of five processing FPGAs (labeled `proc'). 
Each of the latter is driving its associated DAC with the adequate frequency comb which can feature up to 80 tones. The `proc' FPGA is connected to the `split' FPGA with two links. The first of these, labeled `fake ADC', is a 12 bit parallel LVDS link running at 250\,MSPS that is carrying the part of bandwidth corresponding to the excitation signal. The second link, labeled `GTX link', is periodically (at $\sim$953\,Hz) conveying the 80 DDC results over a 2\,Gb/s serial link. The six FPGAs are from the same founder (Xilinx \mbox{XC6VLX75T-2FFG484C}). They provide a satisfactory amount of available user logic, coupled to a sufficiently large Multiplier Accumulator block (MAC) count. They also feature eight high speed serial links.

An additional slow speed DAC, driven by the `split' FPGA, is implemented in order to be able to provide a $\sim$500\,Hz modulation signal. Provided that the board can be clocked with a reference clock, a bidirectional port was provided to allow synchronization between several boards performing acquisition on the same kilo-pixel camera. When using several NIKEL electronics, one board must be configured as master and provide the synchronization signal, while the others are configured as slaves and should start their acquisition upon reception of this synchronization signal.

The communication with the hardware is ensured via a USB2 capable micro-controller and an interface FPGA that accommodates the different voltage levels. It allows the dynamic FPGA reconfiguration, the tone frequencies adjustments and the data readout.

\begin{figure}[th]
\begin{center}
\includegraphics[angle=0,width=0.8\textwidth]{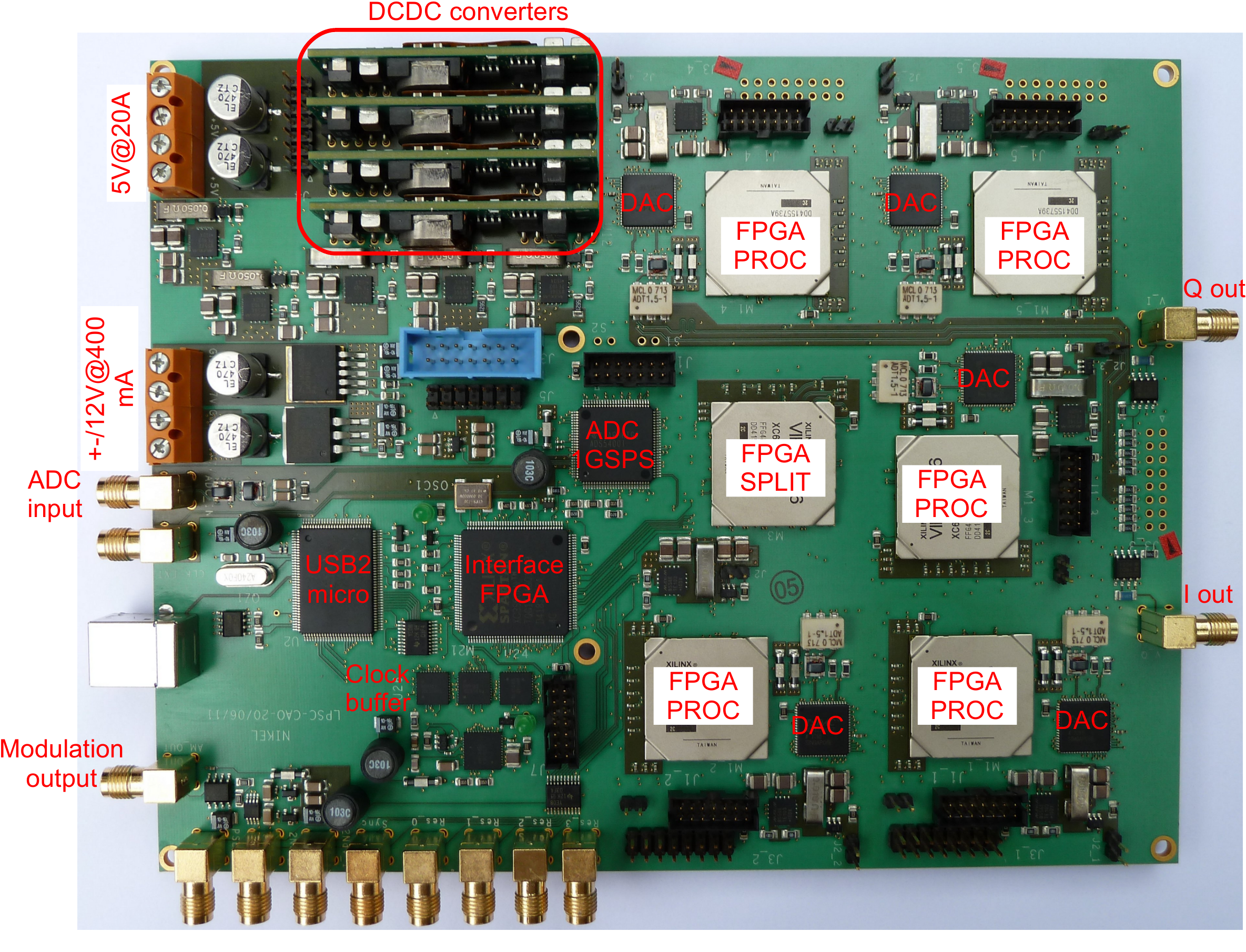}
\caption{Picture of the NIKEL board.  It is a 14 layers PCB having a dimension of $\rm 184\ mm  \times 153\ mm$.}
\label{PictureNikel}
\end{center}
\end{figure}

A picture of the board can be seen in figure~\ref{PictureNikel}. It is a 14 layers PCB having a dimension of $\rm 184\ mm  \times 153\ mm$. The inner dielectric layers are made of traditional FR4 epoxy while the outside layers consist of RO4350 high frequency circuit \cite{ROGERS} that have lower dielectric losses and therefore are well suited to accommodate the 2\,Gb/s serial links, the DAC outputs that provides samples at 1\,GSPS and the ADC input signal.

Due to to the extensive FPGA resource usage and their running frequency (250\,MHz) special care was taken in designing the electronic board power supply. Indeed, each FPGA core supply draws a current of about 5\,A when all tones are activated. Thanks to the usage of DC/DC converters the total current drawn on the input power supply is below 20\,A, thus a maximum required total power of 100\,W (or 0.25\,W per channel).

\section{Polyphase filter design and full chain simulation}
\label{PolyphaseFilterSec}
As introduced in sections~\ref{InstruMetho} and~\ref{HardwareDevel}, the received signal, sampled at 1\,GS/s, must be decomposed in five 250\,MS/s data streams, each stream having a useful bandwidth of 100\,MHz in order to cover the 0-500\,MHz full bandwidth. Frequency modulation/demodulation is a well documented digital signal processing technique \cite{Vaidyanathan} for data transmission (channelization) or audio/image coding application. Those classical techniques, based on Discrete Fourier Transform (DFT), Modified DFT (MDFT) or Cosine Modulated filter banks, suffer from aliasing and data distortion mainly due to the critical sampling of created sub-bands. Overlapping polyphase filter banks as described in \cite{Comoretto}, offer a computationally efficient solution and have only two drawbacks for our application: a first sub-band which is half the bandwidth of the others and the last sub-band (also half bandwidth) is not usable. The same technique can however be adapted to the wanted filter bank specificity, with an acceptable increase of complexity.

\subsection{Theoretical formulation}
\label{theoricSec}
Digital filter banks implementations are often non-intuitive, but are however composed of simple successive digital signal processing blocks, re-arranged in different form to increase computing efficiency. The simplified processing for each band of the filter bank is described hereafter. At first, a frequency shift is performed to translate the band of interest around 0\,Hz. Then a low pass filtering followed by decimation is applied to select the frequencies of interest. This paragraph described the basic blocks arrangement involved in the specific processing used here. 

\begin{figure}[th]
\begin{center}
\includegraphics[angle=0,width=0.95\textwidth]{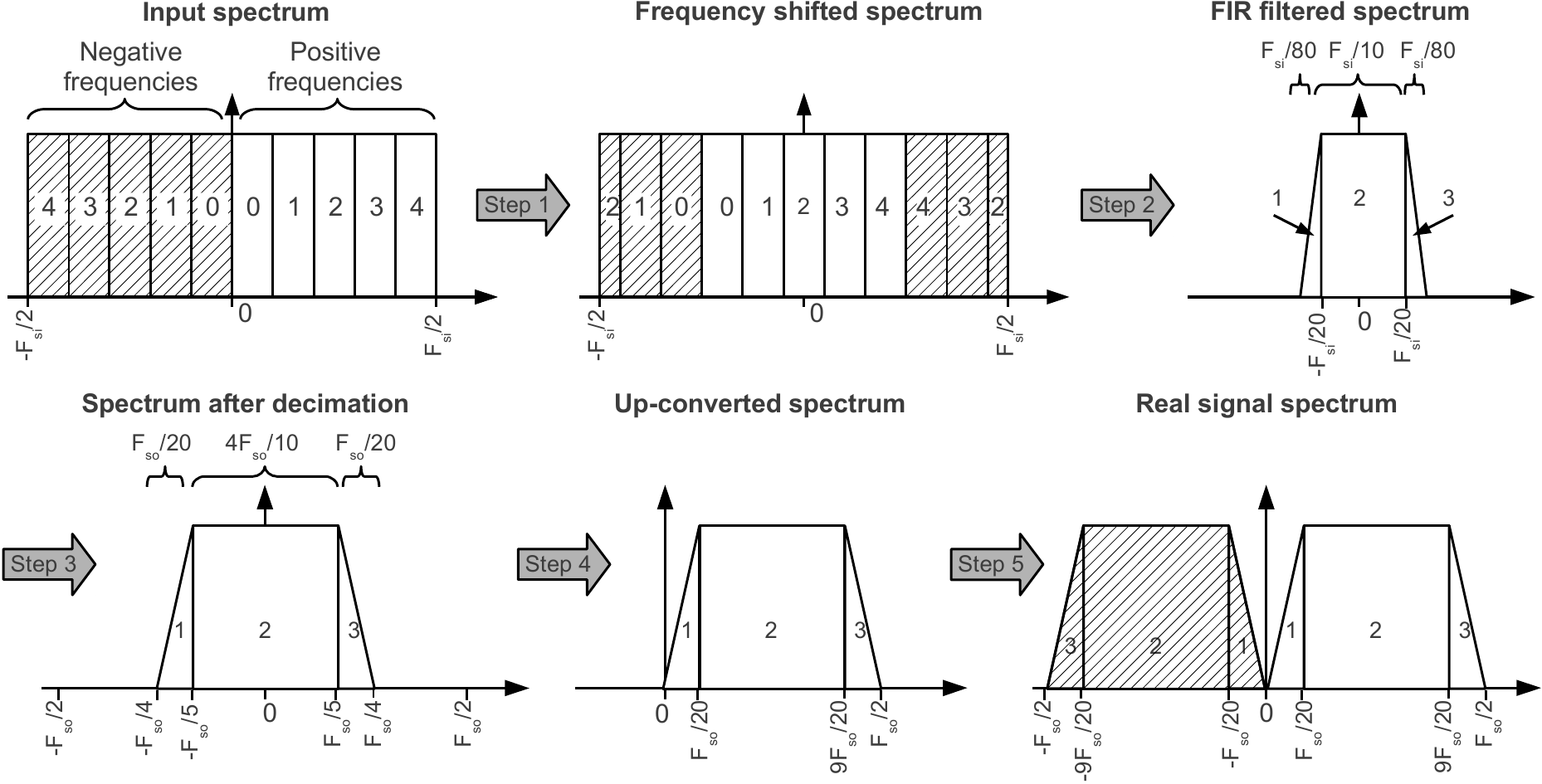}
\caption{Illustration of the polyphase filtering algorithm detailed for band k=2.}
\label{Filter_bank_figure}
\end{center}
\end{figure}

The input data stream is a real signal, sampled at $\rm F_{si}=1\, GS/s$ where four consecutive samples are presented at the filter bank input at each system clock cycle (250\,MHz) while the filter bank outputs five different samples, one for each output band. The signal processing for each band k (k=0..4) is done in five consecutive steps. An illustration is provided for band k=2 in figure~\ref{Filter_bank_figure} and the operations are described hereafter:
\begin{enumerate}
\item Perform an input signal frequency shift of $\rm -(2k+1)\cdot F_{si}/20$  where $\rm F_{si}/20=$50\,MHz. This is obtained by multiplying the input signal by the complex exponential $\rm e^{-j\pi (2k+1)n/10}$ where n is the input signal sample index.

\item Filter the complex signal by a low pass Finite Impulse Response (FIR) filter having a passband of $\rm F_{si}/20$ and a maximum rejection after $\rm F_{si}/10+ F_{si}/80=F_{si}/16$.

\item Decimate the result by a factor of 4. The new data rate then becoming $\rm F_{so} = F_{si}/4$ and the resulting filtered signal bandwidth $\rm [-F_{so}/4,F_{so}/4]$.

\item Up convert by $\rm F_{so}/4$. In practice, realized by multiplying the previous signal by $e^{-j\pi m/2}$ where $m=n/4$ is the sample index of the decimated data stream. The resulting complex signal covers the frequency band $\rm [0, F_{so}/2]$

\item  Finally, keep only the complex signal real part. This will add the complex conjugate negative frequency image in the frequency plane. The output real signal is then correctly sampled at $\rm F_{so}$ without aliasing.
\end{enumerate}

For a given tone c (frequency $\rm f_{ck}$), located in the k\textsuperscript{th} band of the input data stream can have its frequency expressed as $\rm f_{ck} = F_{si}/10 \cdot k + f_c$. Due to the whole processing, it should be noted this tone will not appear in the k\textsuperscript{th} filter bank output at the frequency $\rm f_{c}$, but at $\rm f_c+F_{so}/20$. 
Consequently, the tones used for KID excitation must present a frequency shift of $\rm -F_{so}/20$ with respect to the one used for performing DDC on the returning signal provided by the filter. 

In practice, the FIR filter do not have to be as steep as noted in step 2 given the fact that any aliasing causing frequency folding in the useless sidebands causes no harm. Consequently, a poorer rejection up to $\rm F_{si}/16+ F_{si}/80=3F_{si}/40$ can be tolerated and greatly eases the FIR filter design. 

Unfortunately, this processing is very inefficient in this direct form for several reasons. For instance, the filtering is done for each band and on complex data. Furthermore, resource consuming FIR filtering is performed on the frequency shifted data, but it is followed by decimation. In other words, samples are computed needlessly.
These computing inefficiencies can be considerably improved by grouping the different frequency shifts and by using polyphase filters.

\subsection{Polyphase filters}
If $x(n)$ is the input sample signal, $x^{'}_k(n)$, the frequency shifted data stream for the band k, k=0..4, is expressed by equation~\ref{eq::1}.
\begin{equation}
 x^{'}_k(n)=x(n)\cdot e^{-2j\pi\frac{(2k+1)n}{20}}
\label{eq::1}
\end{equation}

The output, $ x^{''}_k(n)$, of the low pass FIR filter with coefficients $a(p)$ is then
\begin{equation}
 x^{''}_k(n)=\sum_p a(p) \cdot  x^{'}_k(n-p)=e^{-2j\pi\frac{(2k+1)n}{20}} \sum_p a(p)\cdot  x(n-p) e^{2j\pi\frac{(2k+1)p}{20}}
\label{eq::2}
\end{equation}

Provided that the filtered signal is down-sampled by a factor of 4, $ x^{''}_k(n)$ can be only computed for $n=4 m$. By decomposing the filter into a 20 phases polyphase filters, where the  coefficients index p is given by $p = q+20r$ with q=0..19, equation~\ref{eq::2} can be written in the following form
\begin{equation}
 x^{''}_k(m)=e^{-2j\pi\frac{(2k+1)m}{5}} \sum_{q=0}^{19} e^{2j\pi\frac{(2k+1)q}{20}} \cdot w_q(m)
\label{eq::3}
\end{equation}

where $w_q(m)$ is the output of the q\textsuperscript{th} phase polyphase filter.
\begin{equation}
 w_q(m)=\sum_r a(q+20r)\cdot x(4m-q-20r)
\label{eq::4}
\end{equation}

The final step is to up convert the signal by $\rm F_{so}/4$, which is equivalent to a complex multiplication by $\rm j^m$ and taking the real part of the resulting complex number
\begin{equation}
 y_k(m)=  Re \left[e^{j\frac{\pi}{2}m} \cdot e^{-2j\pi\frac{(2k+1)m}{5}} \cdot  \sum_{q=0}^{19} e^{j\pi\frac{(2k+1)q}{10}} \cdot w_q(m)\right]
\label{eq::5}
\end{equation}

The use of the polyphase decomposition of the FIR filter considerably reduces the computation cost. However, it can be seen in equation~\ref{eq::5} that a lot of calculation still need to be done on complex numbers before keeping only the real part. This leaves some margins for optimization.

\subsection{Optimized reconstruction}
\label{OptRecSec}
Since input and outputs of the polyphase filter banks are real signals, it is possible to perform all computation only on real numbers. Equation~\ref{eq::5} can be re-written as
\begin{equation}
 y_k(m)=  Re \left[\sum_{q=0}^{19} e^{j\frac{\pi}{10}[5m+(2k+1)(q-4m)]} \cdot w_q(m)\right]
\label{eq::6}
\end{equation}

We can change the order of the polyphase filter outputs $w_q(m)$ in the sum by introducing new data streams $w^{'}_l(m) =w_q(m)$ with $l = q +( m\  mod\  20)$. Due to the $2j\pi$ periodicity of the complex exponential function, the output of the filter bank can be expressed by the following formula:
\begin{equation}
 y_k(m)=  Re \left[(-1)^{mk} \cdot \sum_{l=0}^{19} e^{j\frac{\pi(2k+1)l}{10}} \cdot w^{'}_l(m)\right]= (-1)^{mk} \cdot \sum_{l=0}^{19} \cos \left(\frac{(2k+1)l\pi}{10}\right) \cdot w^{'}_l(m)
\label{eq::7}
\end{equation}

This simple rotation of the polyphase filter outputs orders, greatly simplify the formula. Moreover, each filter bank output can now be computed without complex arithmetics.

\subsection{Full chain simulation}
\subsubsection{Excitation DAC}
In order to validate the DAC choice and to select its best configuration for each band, that are the digital modulator frequency and the half-band filters to engage, the DAC behavior was simulated.
Indeed, the Frequency Tuning Word (FTW) allowing the configuration of the modulator frequency is given by the following formula $FTW=\dfrac{f_{carrier}}{f_{nco}} \times 2^{32}$, where $\rm f_{nco}$ is 500\,MHz. 
Ideally, it is desired to have five frequency bands and thus five different carrier frequencies going from 0 to 400\,Mhz in steps of 100\,MHz. Consequently, the first approach would be to select these exact values that are perfectly suited to fit with the half-band filters. Unfortunately, these carrier frequencies would yield real FTW instead of integer FTW. Using these rounded values would induce a small offset in frequency which would be observed as a $2 \pi/2^{32}$ phase shift every $2^{32}$ clock cycles. Consequently, the carrier frequencies were adequately chosen to obtain a $\dfrac{f_{carrier}}{500\ MHz}$ ratio of 0, 7/32, 13/32, 19/32 and 26/32 yielding integer FTW values.

Given the fact that the manufacturer provided the DAC half-band filter coefficients, a thorough simulation of the DAC, having the appropriate filters selected, was conducted for the five excitation bands. 
The results confirm that non optimal carrier frequencies are acceptable. In particular, the flatness is slightly degraded while the ripple remains below 0.06\,dB.

This mandatory carrier frequency shift with respect to the ideal value, must be pre-compensated in the FPGA `proc' building the excitation frequency comb by a digital modulator that apply a frequency shift in the opposite direction to virtually obtain carrier frequencies at the requested values (from 0 to 400\,MHz). The required compensations, expressed as a ratio of $\rm F_{so}$, are respectively: 0, -3/80, -2/80, +2/80 and -4/80.

This shift is accomplished in the meantime as the frequency shift of $\rm -F_{so}/20$ needed to compensate the polyphase filter bank induced shift (see section~\ref{FirmwareProcDevel}). Consequently, the final required compensations, again expressed in ratio of $\rm F_{so}$ are : -1/20, -7/80, -4/80, -3/80 and -3/40. In practice, these are implemented with 80 values sine and cosine table feeding digital modulator. 

\subsubsection{Polyphase filter}
Likewise to the excitation DAC, the polyphase filter was simulated in order to assess its performances and to find the best implementation options matching the FPGA available resources. During the firmware design, the mathematical simulation tool was used to build stimulus files and reference filter output that were used by the VHDL simulation tool to speed up the design and validate the firmware implementation of the filter. 

\begin{figure}[th]
\begin{center}
\includegraphics[angle=0,width=0.48\textwidth]{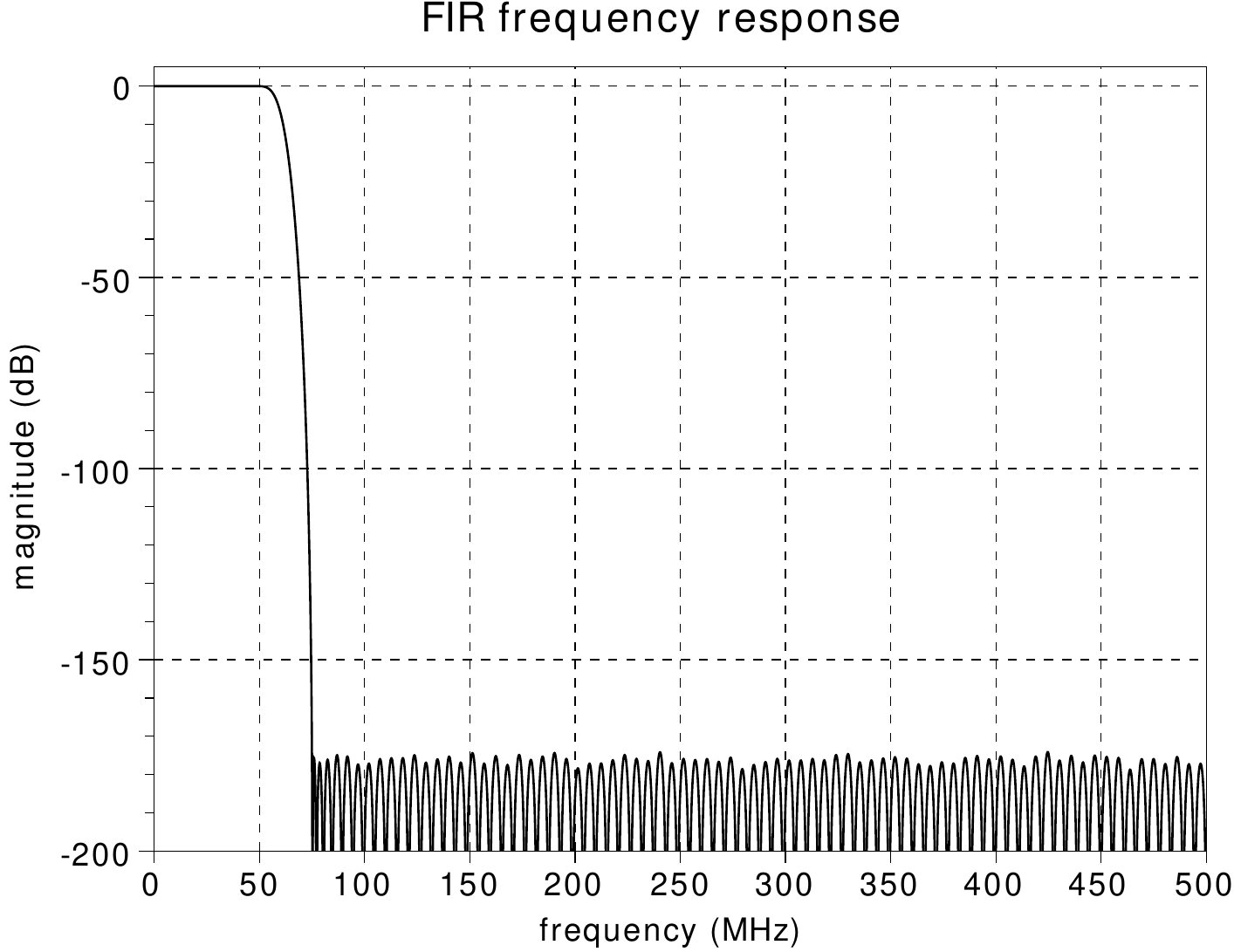}
\includegraphics[angle=0,width=0.48\textwidth]{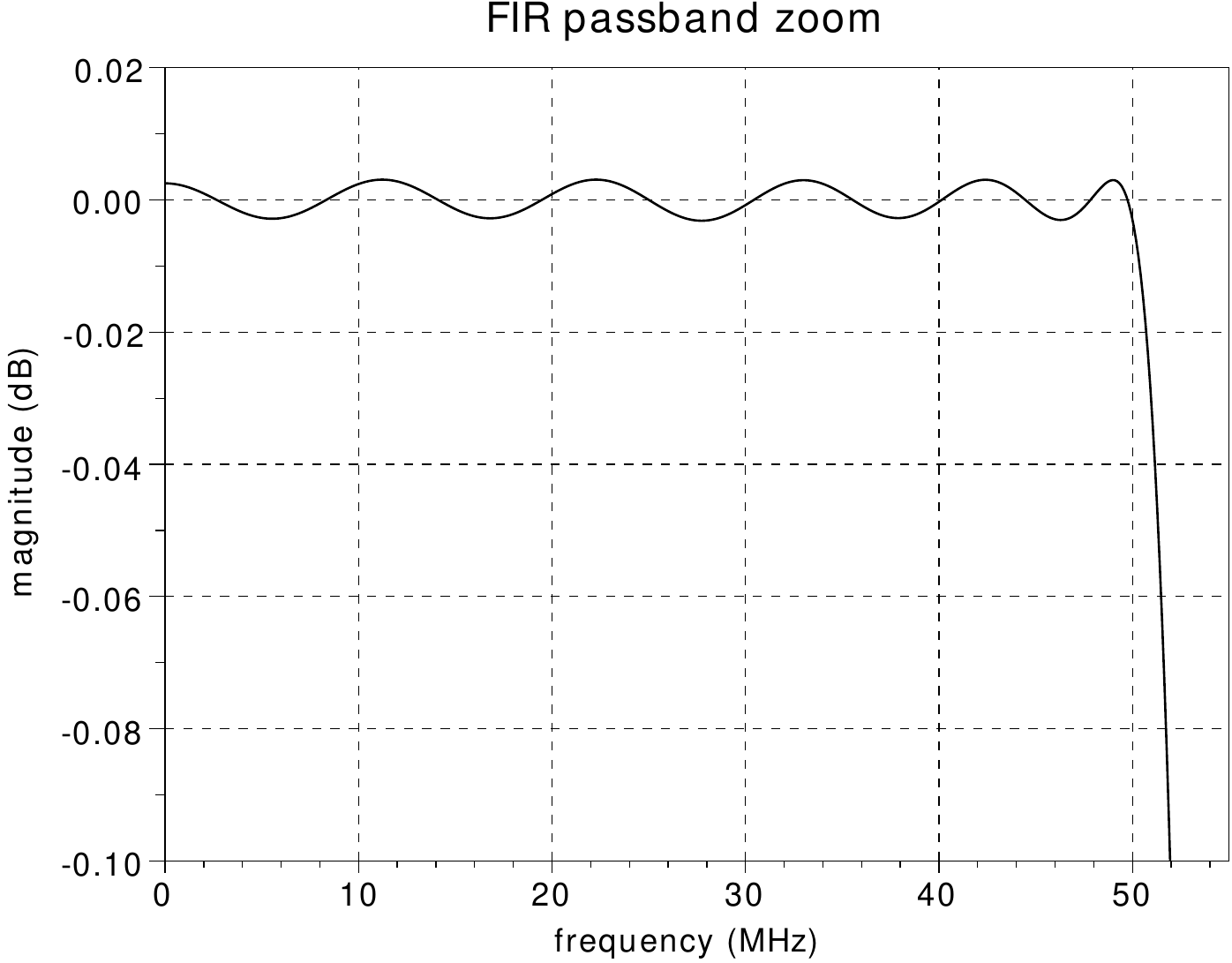}\\
\vspace{10pt}
\includegraphics[angle=0,width=0.48\textwidth]{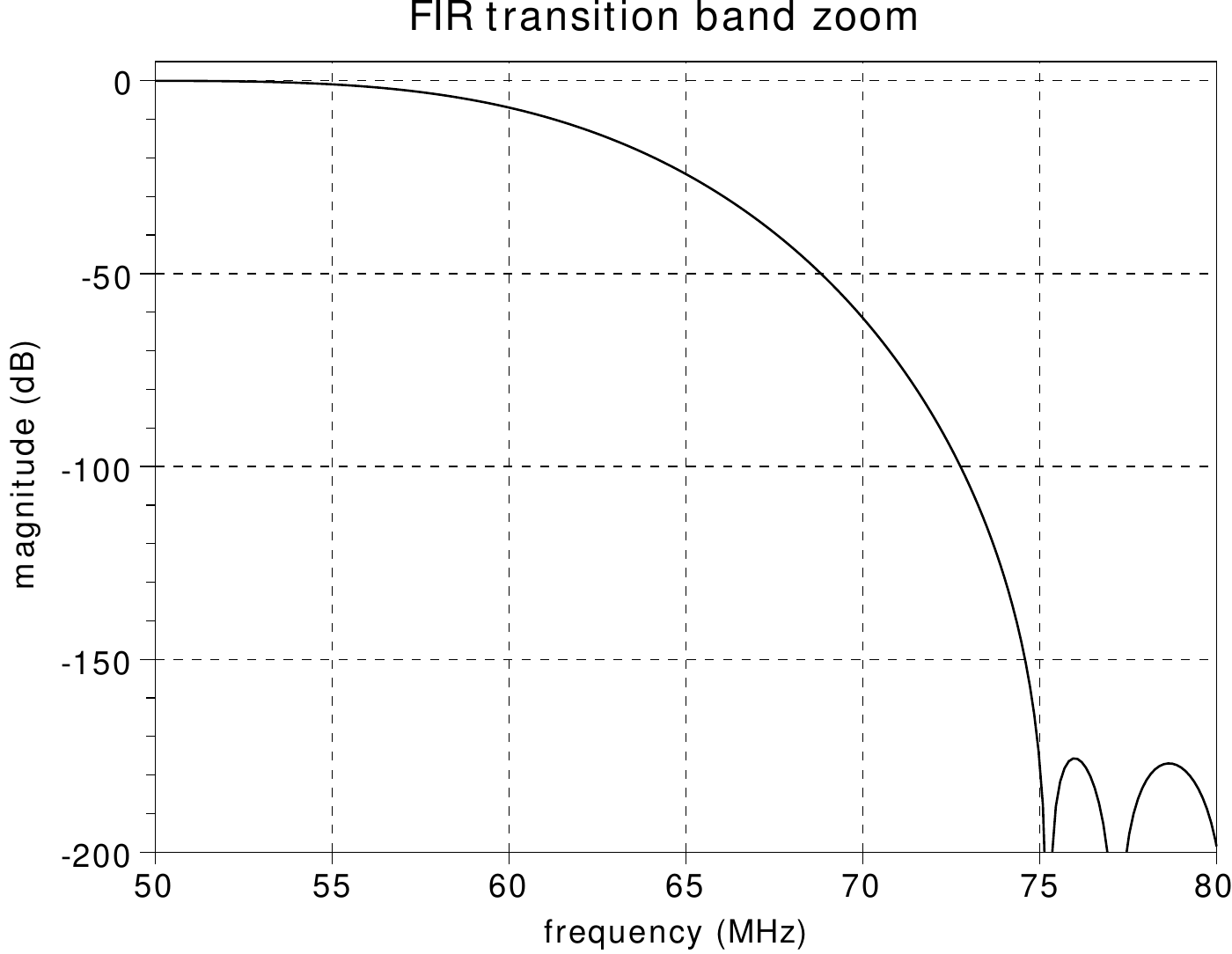}

\caption{Simulation of the selected FIR filter. Top left figure shows the global filter response. Top right figure, shows the gain fluctuation in the passband and bottom figure shows the steep rejection after the passband.}
\label{plot_filtre_nika}
\end{center}
\end{figure}

The simulation was also an asset in designing the FIR filter used. As shown in figure~\ref{plot_filtre_nika}, the selected FIR has a good flatness over the useful bandwidth ($<$0.01\,dB). The choice was made to concede a larger than specified transition band [50-75\,MHz]  while having an excellent rejection (-170\,dB) in the stopband. As explained in section~\ref{theoricSec}, possible resulting aliasing does not impact DDC performances in the useful bandwidth.
Additionally, the quantization noise due to the use of the fixed point Multiplier ACcumulator (MAC) was evaluated and confirmed to be negligible with respect to the quantization noise of the ADC.

\begin{figure}[th]
\begin{center}
\includegraphics[width=0.48\textwidth]{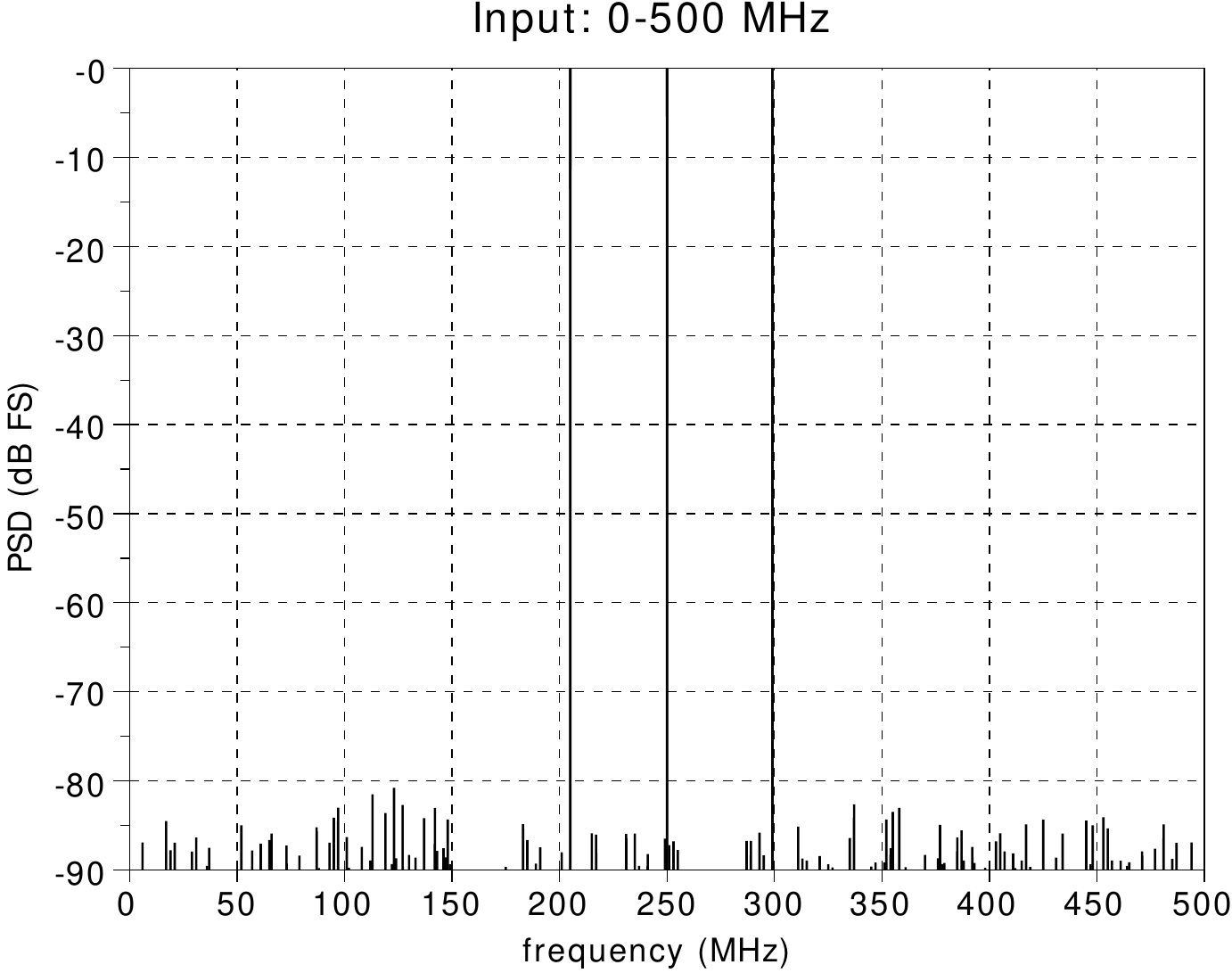}
\includegraphics[width=0.48\textwidth]{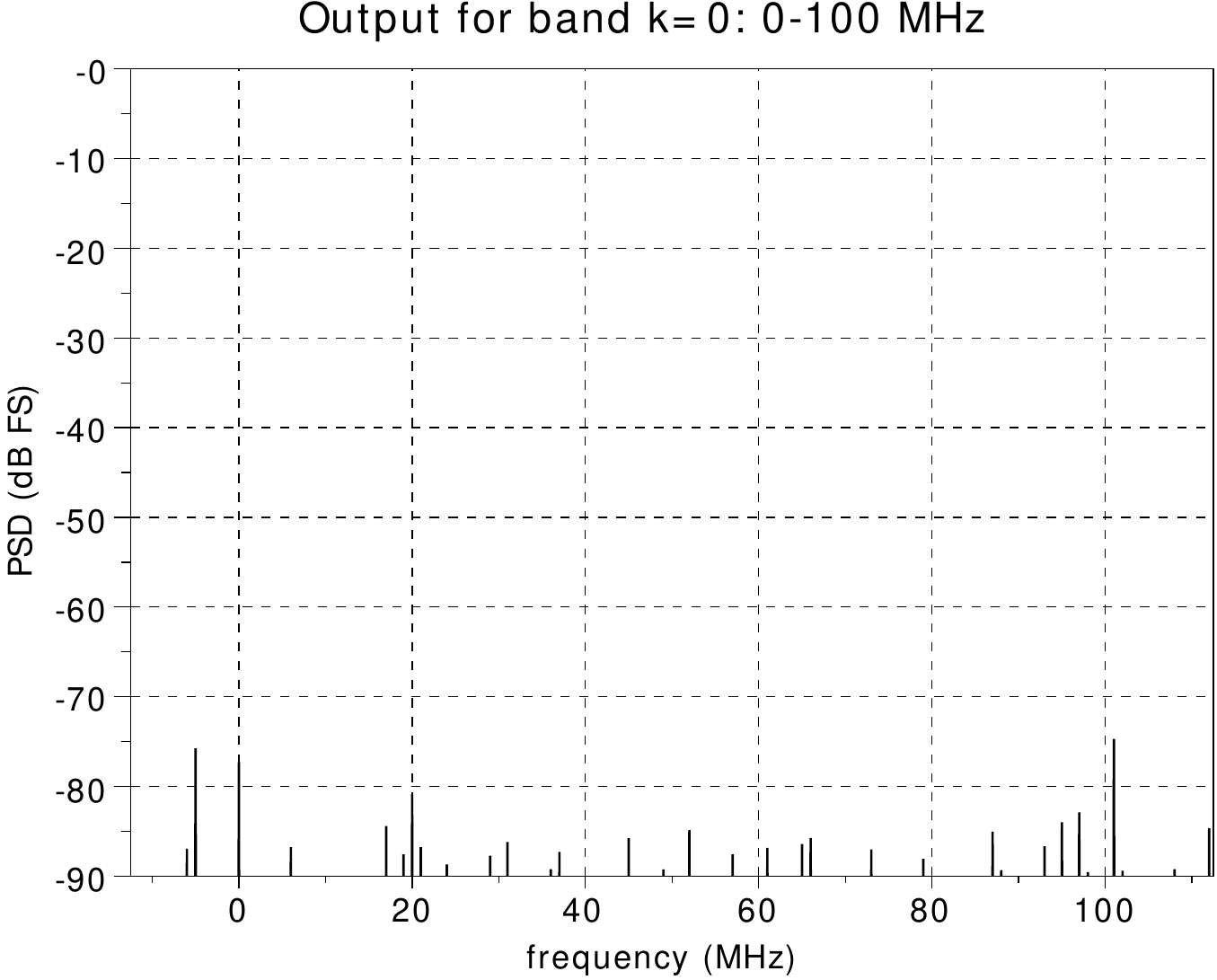} \\
\vspace{10pt}
\includegraphics[width=0.48\textwidth]{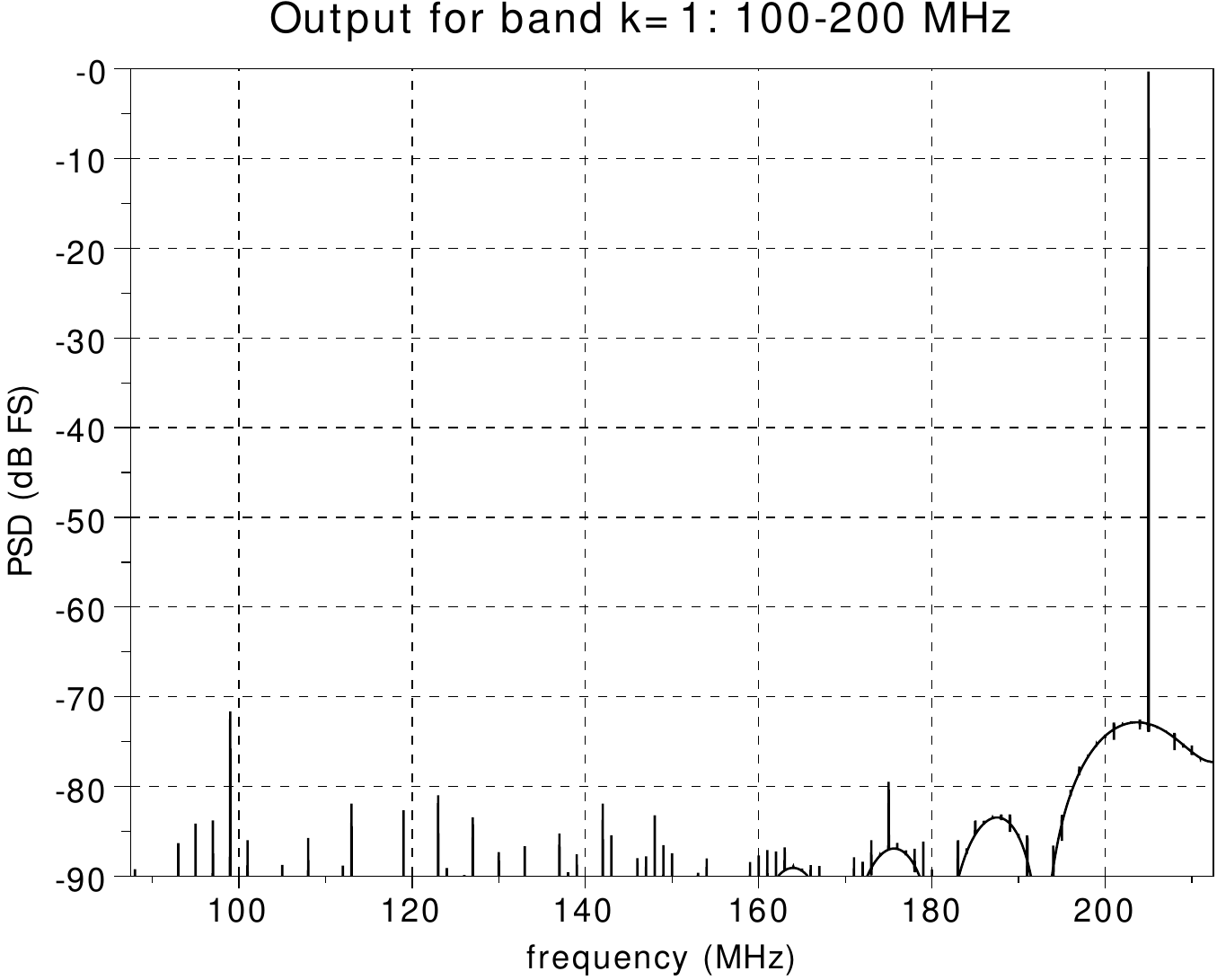}
\includegraphics[width=0.48\textwidth]{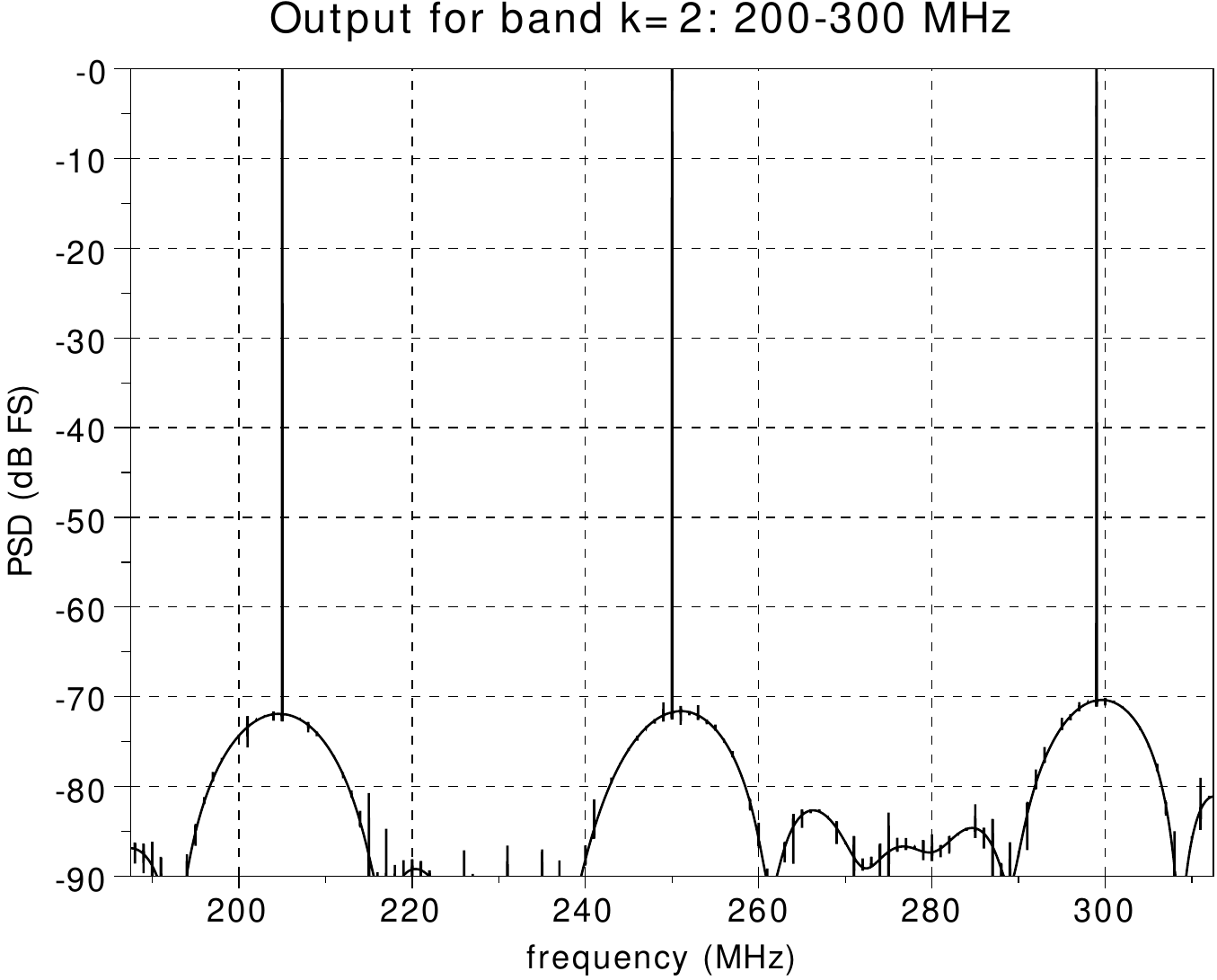}\\
\vspace{10pt}
\includegraphics[width=0.48\textwidth]{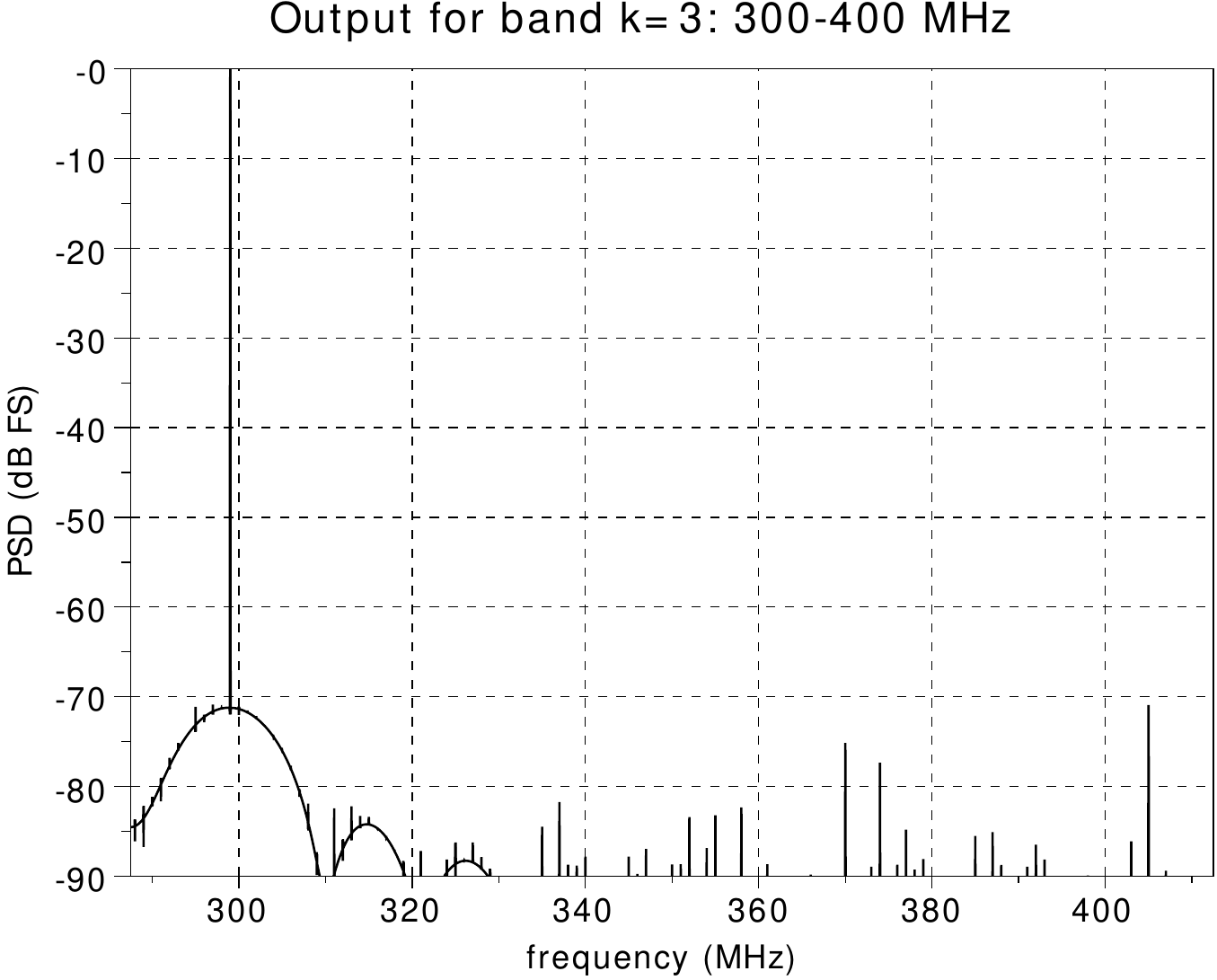}
\includegraphics[width=0.48\textwidth]{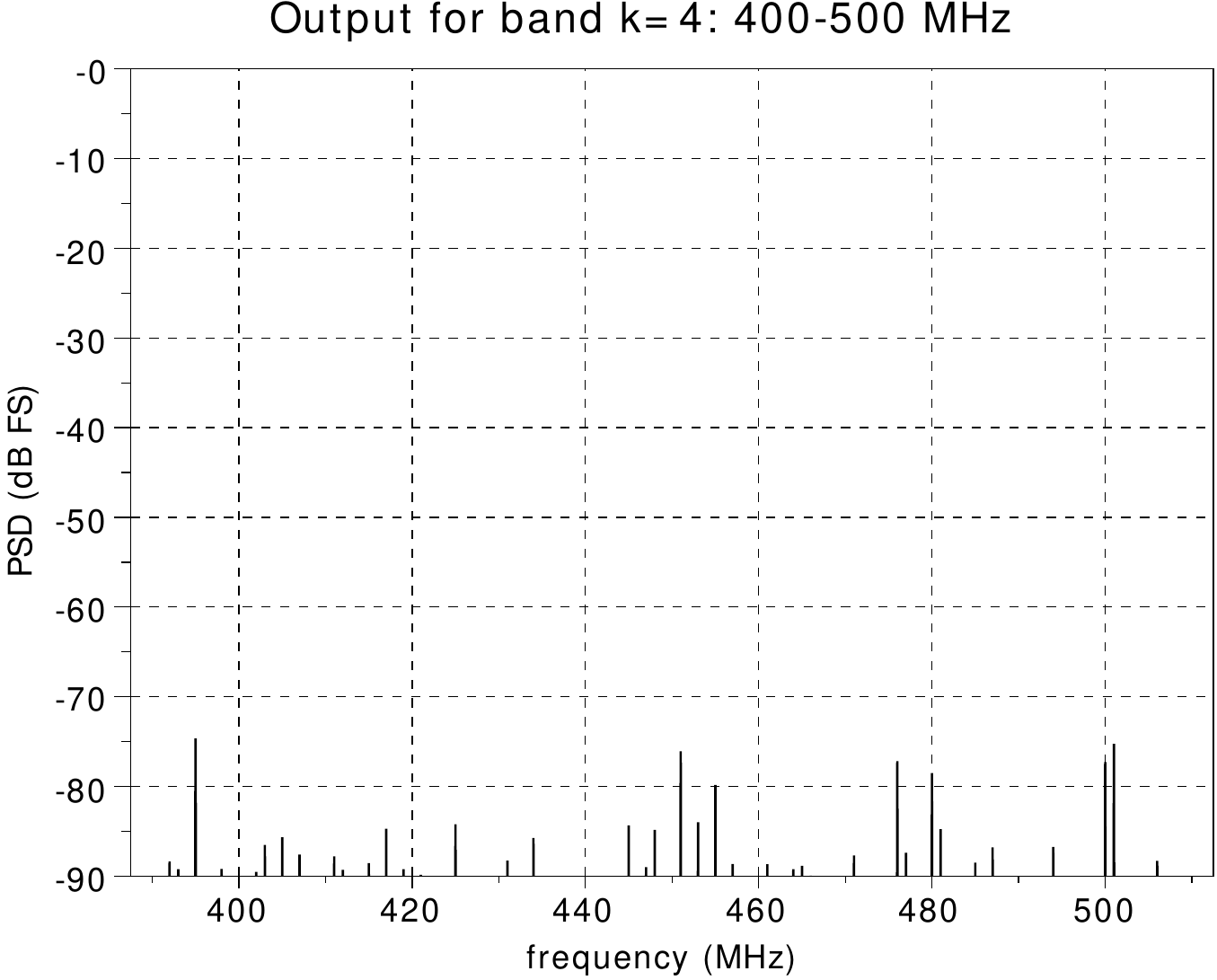}

\caption{full polyphase filter simulation, where three tones (205\,MHz, 250\,MHz, 299\,MHz) are injected at the filter input. The top left figure shows the input signal spectrum, and the other plots show the frequency content of each output of the polyphase filter. It can be observed that the expected tones lie in the expected band k=2, while the spurious appearing in band k=1 and k=3 are in their rejected side bands, i.e above 200\,MHz for band k=1 and below 300\,MHz for band k=3.}
\label{plot_decompositionnika}
\end{center}
\end{figure}

A full polyphase filter simulation, where three tones (205\,MHz, 250\,MHz, 299\,MHz) are injected at the filter input, is shown in figure~\ref{plot_decompositionnika}. The top left figure shows the input signal spectrum, and the other plots show the frequency content of each output of the polyphase filter. It can be observed that the expected tones lie in the expected band k=2, while the spurious appearing in band k=1 and k=3 are in their rejected side bands, i.e above 200\,MHz for band k=1 and below 300\,MHz for band k=3.

\section{Firmware development}
\label{FirmwareDevel}
\subsection{FPGA `split' description}
\label{FirmwareSplitDevel}
\begin{figure}[th]
\begin{center}
\includegraphics[angle=0,width=0.95\textwidth]{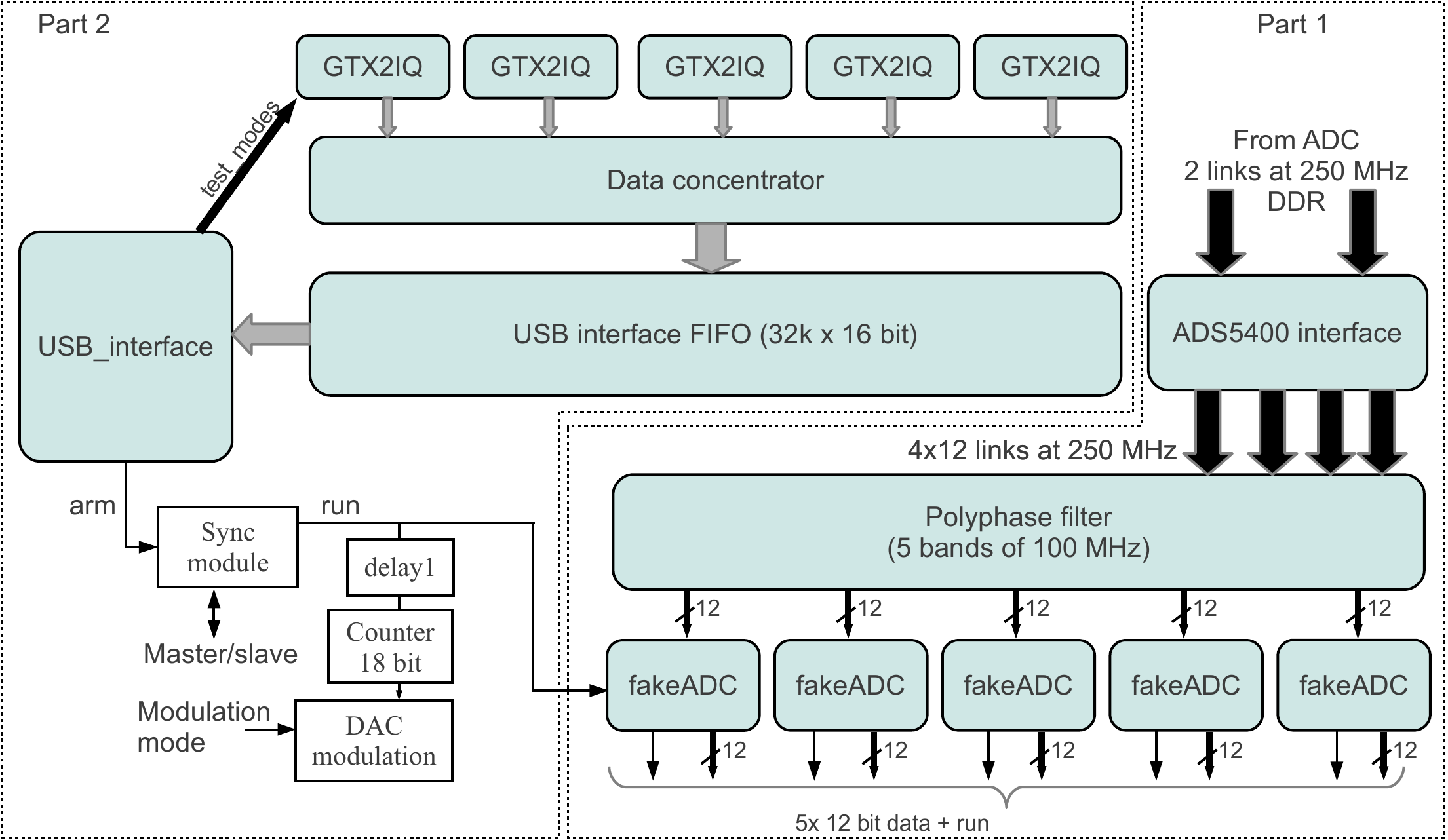}
\caption{Overview of the `split' FPGA firmware. The firmware is divided in two main parts. The first part, which is the key-point of the overall design, is composed of the ADC interface, the polyphase filter bank and the five `fakeADC' outputs, each carrying its share of the bandwidth to the dedicated `proc' FPGA. The second part consists of five GTX receivers that collect the I/Q data provided by each `proc' FPGA, a data concentrator, a large FIFO and a USB interface.}
\label{FPGA_split_nikel}
\end{center}
\end{figure}

The `split' FPGA, shown in figure~\ref{FPGA_split_nikel}, contains two main parts. The first part, which is the key-point of the overall design, is composed of the ADC interface, the polyphase filter bank and the five `fakeADC' outputs, each carrying its share of the bandwidth to the dedicated `proc' FPGA. The second part consists of five GTX receivers that collect the I/Q data provided by each `proc' FPGA, a data concentrator, a large FIFO and a USB interface.

The GTX2IQ receiver blocks are designed to operate at a speed of 2\,Gb/s. This is the speed required to carry 32 bit at 50\,MHz with an 8b/10b encoding. 
Every $\sim$1.05\,ms (2\textsuperscript{18} clock cycles at 250\,MHz) a 644 bytes data frame is received (see section~\ref{FirmwareProcDevel}) and stored in a small reception buffer (1\,k word deep). 
Once all GTX2IQ received its data frame, the `data concentrator' transfers each link data into the global data buffer labeled `USB interface FIFO' (32\,k word deep) to make the complete data frame available for data acquisition via the USB interface.

The USB interface is mostly in charge of reading out the `USB interface FIFO' and thus of performing data acquisition. The required data throughput is $\rm 644 \times 5 \times 953=3\ MB/s$. The interface is also used to set the master/slave mode, to arm the acquisition, to select the modulation mode and to configure and recover the status of the GTX transceiver links.

The DAC modulation block is used to generate an optional modulation signal which can be a 2 or 4 values modulation signal, depending whether it is desired or not to compute the sensitivity (first derivative) and the sensitivity variation (second derivative) of the I/Q measurement. When this block is activated, the modulation signal is modified every integration cycle. To ensure the modulation synchronousness with the integration performed in the DDC, the initial start of the modulation is adjustable with a resolution of 4\,ns and up to one full integration cycle. 

\begin{figure}[th]
\begin{center}
\includegraphics[angle=0,width=0.7\textwidth]{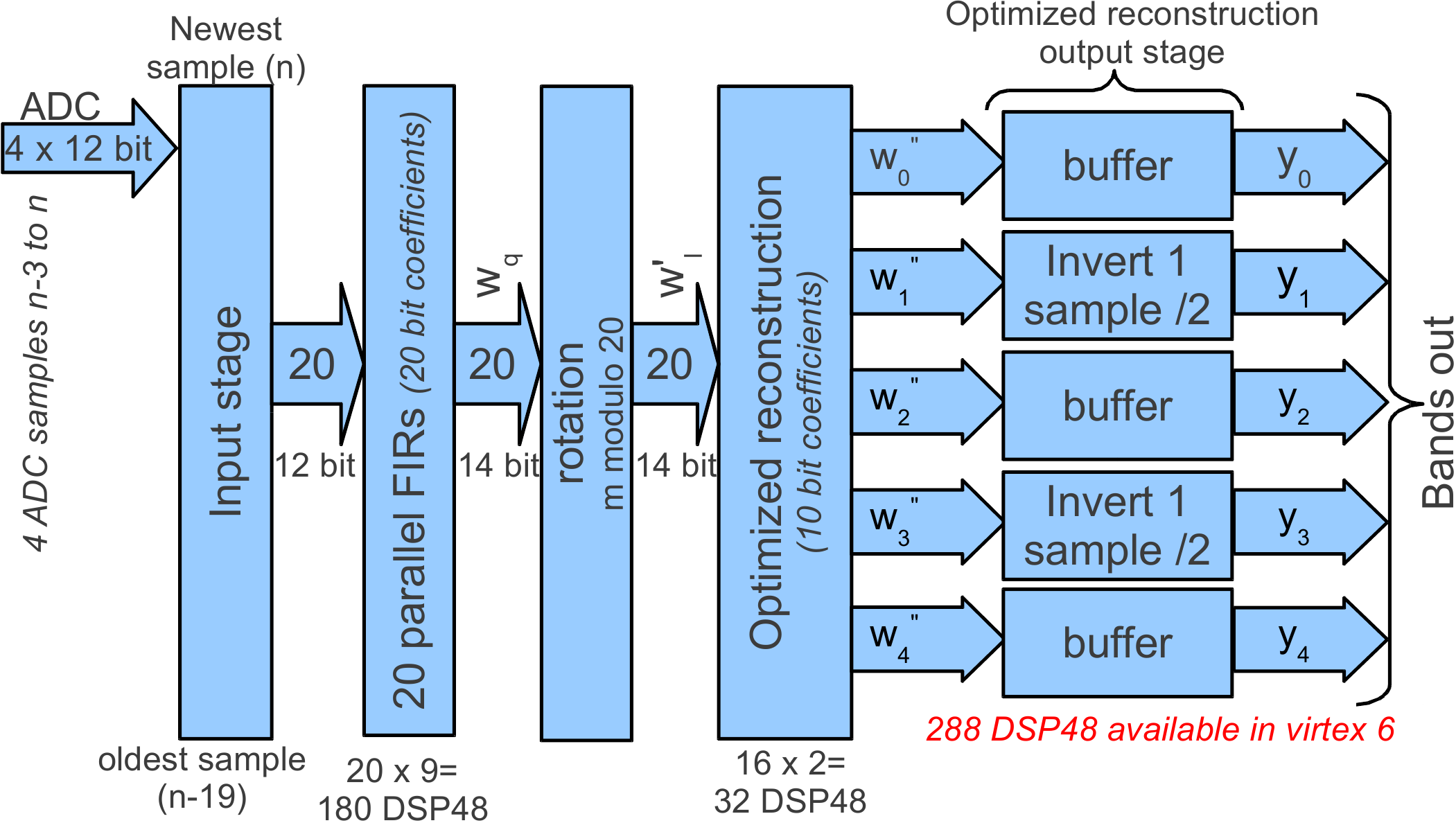}
\caption{Polyphase filter bank implementation overview.}
\label{0_ArchiFiltreSeparation}
\end{center}
\end{figure}

The polyphase filter bank implementation (shown in figure~\ref{0_ArchiFiltreSeparation}) is composed of five successive stages. During the design, several stratagems where used to minimize the number of DSP48 blocks used and hence to allow the filter to fit in the chosen FPGA.

The input stage is composed of a shift register bank featuring 20 registers of 12 bit. It receives four new ADC samples every clock cycle and at the same time performs a four data samples shifting from the newest data to the oldest. At the output of this stage the $\rm n$ to $\rm n-19$ samples are provided to the following stage.

The following stage is composed of 20 FIR filters, each processing one of the `input stage' output. The FIR filters feature 45 taps and are implemented in the transposed direct form which suits perfectly the possibilities offered by the DSP48 blocks inside the Virtex 6 FPGA. Given the fact that for each FIR filter only 9 taps out of 45 are non zero, the zero coefficient taps are replaced by simple registers. This artifice alone allows to use only 180 DSP48 blocks for the whole filter bank instead of 450.

The third stage, named  `rotation block', is used to rotate the vector composed of the 20 FIR filter outputs and to provide it to  the `optimized reconstruction block'. The rotation consist in routing the data according to the following equation:
 $W^{'}_l(m) =W_{(l- m)\  mod\  20}(m)$ where l=0..19 and m the sample index. 
In practice, this is implemented with 20 high performance multiplexers having 20 inputs and one output. Each of these multiplexers is controlled by a counter having a  0 to 19 range and is initialized with a value according to the `optimized reconstruction block' input it is connected to.

According to equation~\ref{eq::7} given in section~\ref{OptRecSec}, having the vectors Y(m), $\rm  W^{''}(m)$ and $\rm W^{'}(m)$ being respectively composed of $y_k(m)$, $w^{''}_k(m)$ and $w^{'}_l(m)$ for k=0..4 and l=0..19, the optimized reconstruction can be computed by 
$\rm Y(m)=J^{m} \cdot W^{''}(m)=J^{m} \cdot A \cdot W^{'}(m)$, 
where:
\begin{equation} 
J=
\begin{bmatrix} 
1 & 0  & 0  & 0  & 0 \\
0 & -1  & 0  & 0  & 0 \\
0 & 0  & 1  & 0  & 0 \\
0 & 0  & 0  & -1  & 0 \\
0 & 0  & 0  & 0  & 1 \\
\end{bmatrix}
\end{equation}

and 
\begin{equation} 
\setcounter{MaxMatrixCols}{20}
A=
\begin{bmatrix} 
1 & a  & b  & c  & d  &  0 & -d & -c & -b & -a & -1 & -a & -b & -c & -d & 0 & d  & c  & b  & a \\
1 & c  & -d & -a & -b &  0 & b  & a  & d  & -c & -1 & -c & d  & a  & b  & 0 & -b & -a & -d & c \\
1 & 0  & -1 &  0 &  1 &  0 & -1 & 0  & 1  & 0 & -1 &  0 & 1  & 0  & -1 & 0 &  1 &  0 & -1 & 0 \\
1 & -c & -d &  a & -b &  0 &  b & -a & d & c & -1  & c  & d  & -a & b  & 0 & -b & a & -d & -c \\
1 & -a & b  & -c & d  &  0 & -d & c & -b & a & -1  & a  & -b & c & -d & 0 & d & -c & b & -a
\end{bmatrix}
\end{equation}
with:
\begin{equation} 
a=\cos\left(\frac{\pi}{10}\right), b=\cos \left(\frac{\pi}{5}\right),c=\cos\left(\frac{3\pi}{10}\right), d=\cos\left(\frac{2\pi}{5}\right)
\end{equation}

\begin{figure}[th]
\begin{center}
\includegraphics[angle=-90,width=0.9\textwidth]{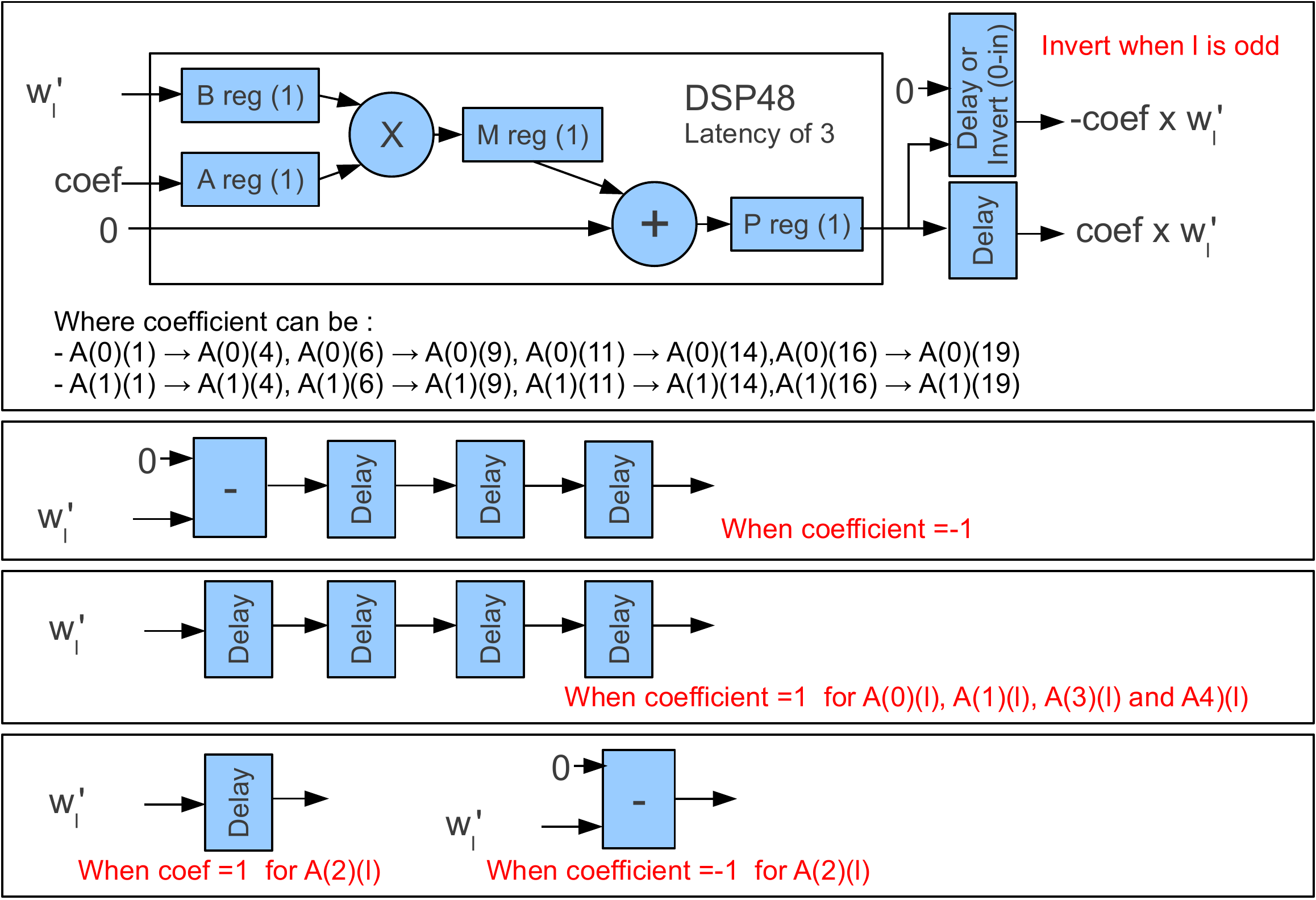}
\caption{Optimized operator taking benefit of the sign symmetry in A matrix between row k=0 and k=4 and row k=1 and k=3. For row k=2 no multiplier is needed and since there is no phase requirement between the different frequency bands, the delay adjustment needed to compensate the DSP48 latency are in fact unnecessary.}
\label{4a_optimized_reconstruction_operator}
\end{center}
\end{figure}

It can be seen that computing $\rm W^{''}(2)$ does not need any multiplier since the sign inversion can be simply obtained by computing the two complement of the input value. Moreover, by using $2 \times 16$ DSP48 slices for computing the non zero and non one values multiplication of the two first row of the A matrix, the last two rows can be obtained by sign inversion only. The sign inversion is applied on one out of two multiplications only (when l is odd). 
Figure~\ref{4a_optimized_reconstruction_operator} provides a visual summary of the block implementation scheme. 
The whole processing must be pipelined and as opposed to a FIR filter, each single sample is multiplied by the 20 coefficients of each row and then the operation results are all summed together. This requires the use of two pipelined adders types: one having ten inputs for $\rm W^{''}(2)$ (with four pipeline stages) and another having 18 inputs for the others (with five pipeline levels). Finally, for a $[5,20] \times [20,1]$ matrix multiplication, only 16 DSP48 slices are used.

The last stage is actually associated with the previous stage  (`optimized reconstruction output stage'), but for the sake clarity it is shown as separate block.
It corresponds to the first term of  equation~\ref{eq::7}, which performs an alternate sign inversion for the odd bands resulting in a frequency shift by half the sampling frequency and in a frequency scale reversion.

The whole design uses 216 out of 288 DSP48 blocks, 18442 out of 93120 slice registers and 14879 out of 46560 slice LUT.

\subsection{FPGA `proc' description}
\label{FirmwareProcDevel}
\begin{figure}[th]
\begin{center}
\includegraphics[angle=-90,width=0.95\textwidth]{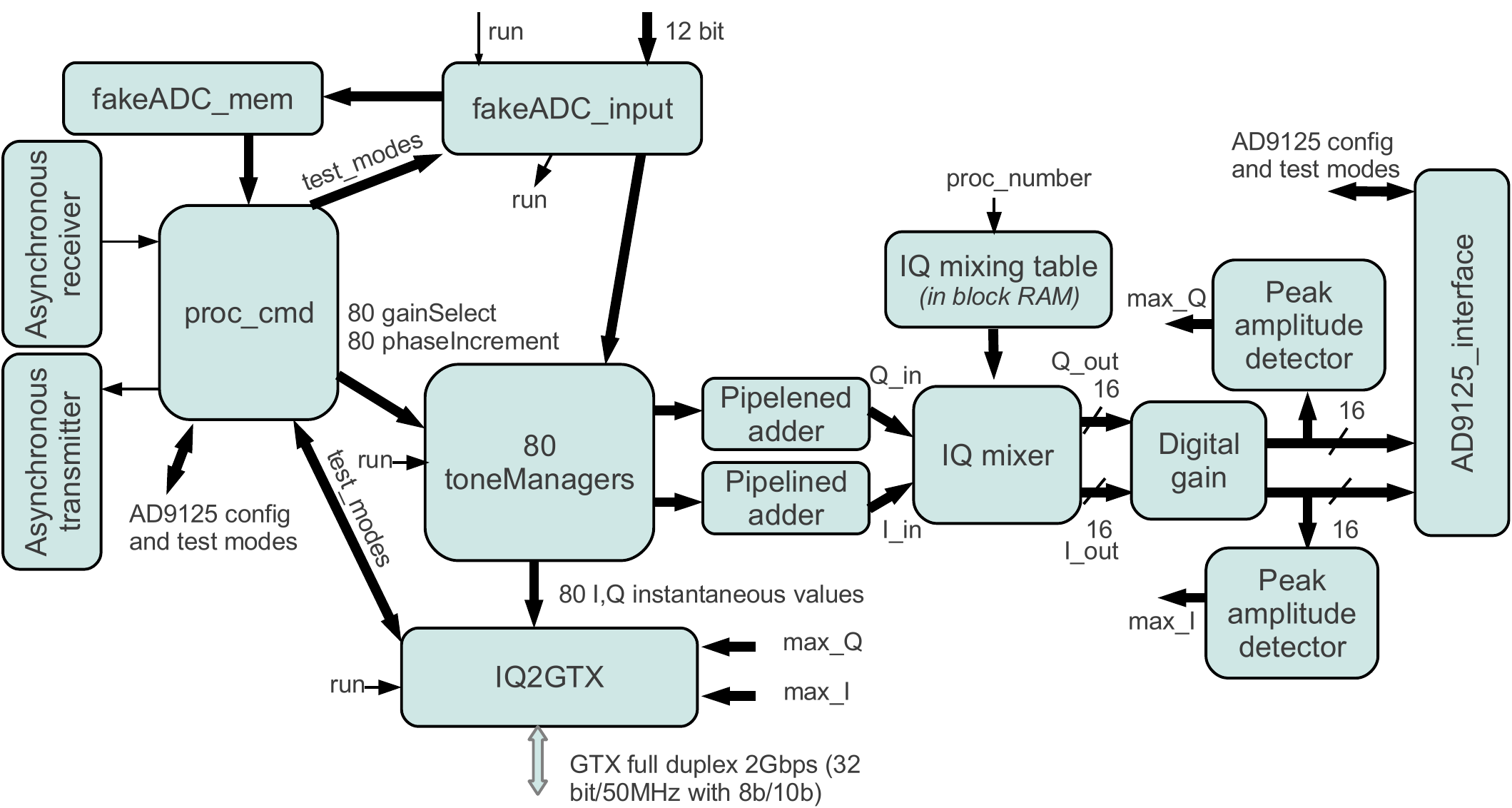}
\caption{Overview of the `split' FPGA firmware. }
\label{FPGA_proc_nikel}
\end{center}
\end{figure}

As explained before, the processing FPGA, whose block diagram is shown in figure~\ref{FPGA_proc_nikel}, is in charge of generating the frequency comb in its share of bandwidth and to perform the channelized DDC for each considered tone.

The communication between the USB interface and the FPGA is ensured via a serial link running at 50\,MHz. The various commands received are interpreted by the `proc\_cmd' state machine. Commands are of two kinds: the write commands and the read commands. The write commands are used, for instance, to set the individual phase increment values and tone attenuation, the digital gain and the mixing table to use for performing a frequency shift. 
Configuration and test modes can also be set via this interface. Among the provided test modes, it may be noted that it is possible to record a  `fakeADC' signal snapshot of 32\,k samples in the `fakeADC\_mem'. The read command are used to request data from the FPGA like the GTX link status, the `fakeADC' link synchronization status. Moreover the DAC internal registers values can be accessed.

Given the fact that the `fakeADC' data emitted by the `split' FPGA are synchronized by the system wide reference clock, a dedicated interface (fakeADC\_input) is used to adjust the `fakeADC' bus delay in order to compensate the data sampling phase misalignment and thus to guarantee stable information sampling. The locally synchronized data are provided to the tone managers.

The 80 tone manager outputs are fed to two pipelined adder in order to construct the in-phase and quadrature versions of the frequency comb. Each comb version is then frequency shifted by an IQ mixer in order to compensate the residual up converting due to the polyphase filtering and the frequency shift due to the non optimal selection of the DAC internal modulator frequency (see section~\ref{PolyphaseFilterSec}). The digital gain is used to numerically amplify the resulting signal by 0\,dB up to 36\,dB in steps of 6\,dB before driving the DAC. This feature is useful to adapt the signal to the ADC input range when less than 80 tones are used.

The IQ2GTX block is used to transmit the DDC results through the high speed link to the `split' FPGA for data concentration. Along with these data, the detected peak amplitude, in absolute value, is transmitted for monitoring and to avoid DAC clipping. Hence, the data frame is composed of $\rm 2 \times 80$ 32 bit words representing the in-phase/quadrature information.

\begin{figure}[th]
\begin{center}
\includegraphics[angle=0,width=0.75\textwidth]{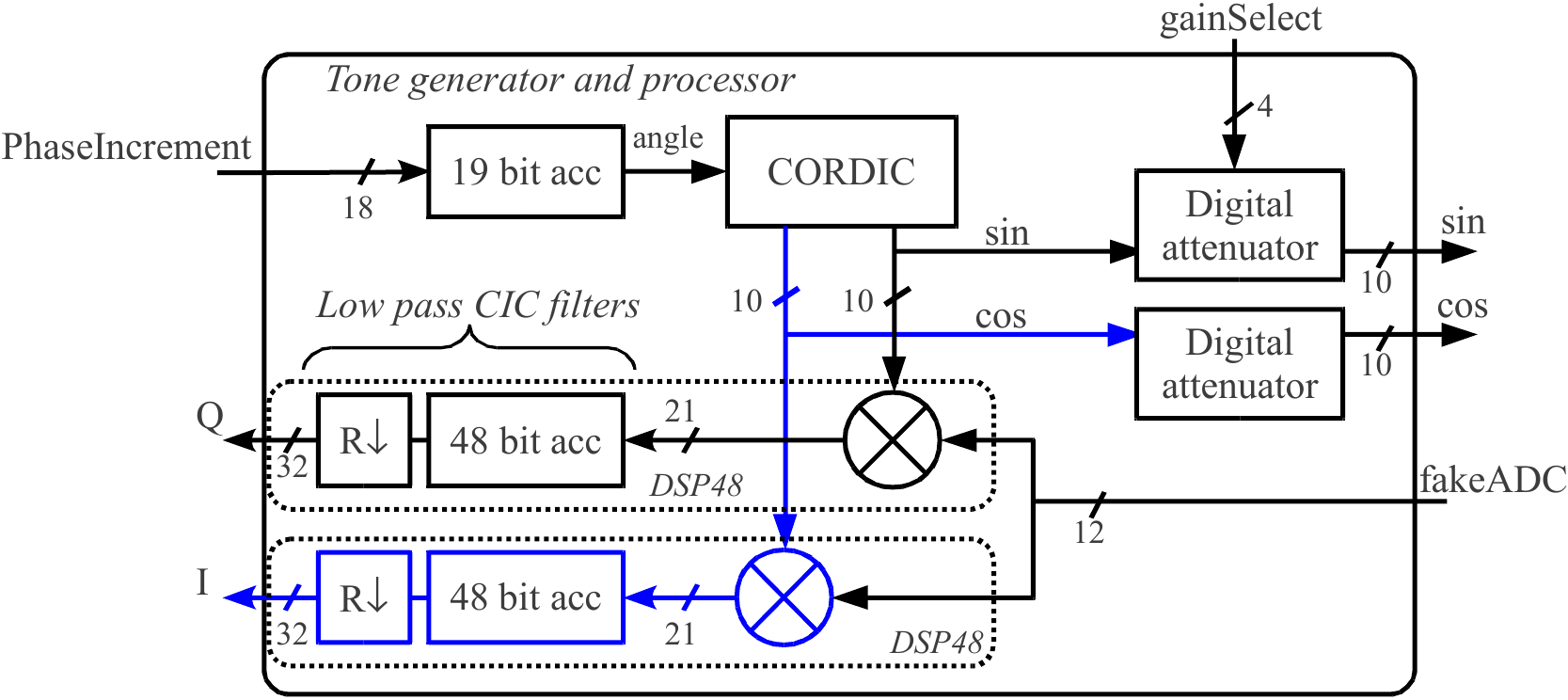}
\caption{Overview of a tone manager. The block comprises a CORDIC generator, two digital attenuators for individual tone power adjustment and a DDC implemented with DSP48 blocks.}
\label{nikel_tone_manager}
\end{center}
\end{figure}

The tone manager, which is depicted in figure~\ref{nikel_tone_manager}, features a COordinate Rotation DIgital Computer (CORDIC) \cite{Volder} block and a DDC that is composed of an I\&Q demodulator followed by a Low Pass Filter (LPF). The LPF, which is primarily used to remove the summed frequencies component from the spectrum, also provides unwanted frequencies rejection (e.g. frequencies tuned to other pixels, white noise, \ldots). Each CORDIC, implemented in a pipelined fashion and composed only of adders and subtracters, was designed to provide a 10 bit precision on the sine and cosine values calculated. It uses 10 precalculated arc tangent values with 20 bit resolution. The phase accumulator that feeds the CORDIC is used to adjust the frequency with a precision of $\rm 250\ MHz/2^{18} \sim 953\ Hz$. In order to avoid in phase startup at the maximum cosine or sine amplitude of all CORDIC, the phase accumulator is initialized at a quarter of its full scale, i.e each phase accumulator is reseted at $\pi/
4$.

The I\&Q demodulation is performed by multiplying a copy of the ADC output by replicas of the generated sine and cosine values. For practical reasons (FPGA logic resources), the Low pass Filter (LPF) is obtained by averaging $2^{18}$ data samples and it is thus in the order of the kHz of bandwidth. It must be noted, that the accumulator period must be chosen as a multiple of the phase accumulator period in order to avoid beat frequency phenomena. At the end of the accumulation cycle, each tone manager transfers its I\&Q data to the IQ2GTX interface for transmission to the `split' FPGA.

To allow individual tone power adjustment, the sine and cosine wave are passed through digital attenuators before being provided to the block output. Tones can be tuned in the range 0 to 8/8 and have a resolution of 1/8\textsuperscript{th} of the input power. 

The whole design uses 164 out of 288 DSP48 blocks, 60412 out of 93120 slice registers and 43508 out of 46560 slice LUT.

\section{Prototype performance}
\label{ProtoPerf}
\subsection{System frequency response}
\begin{figure}[th]
\begin{center}
\includegraphics[angle=0,width=0.7\textwidth]{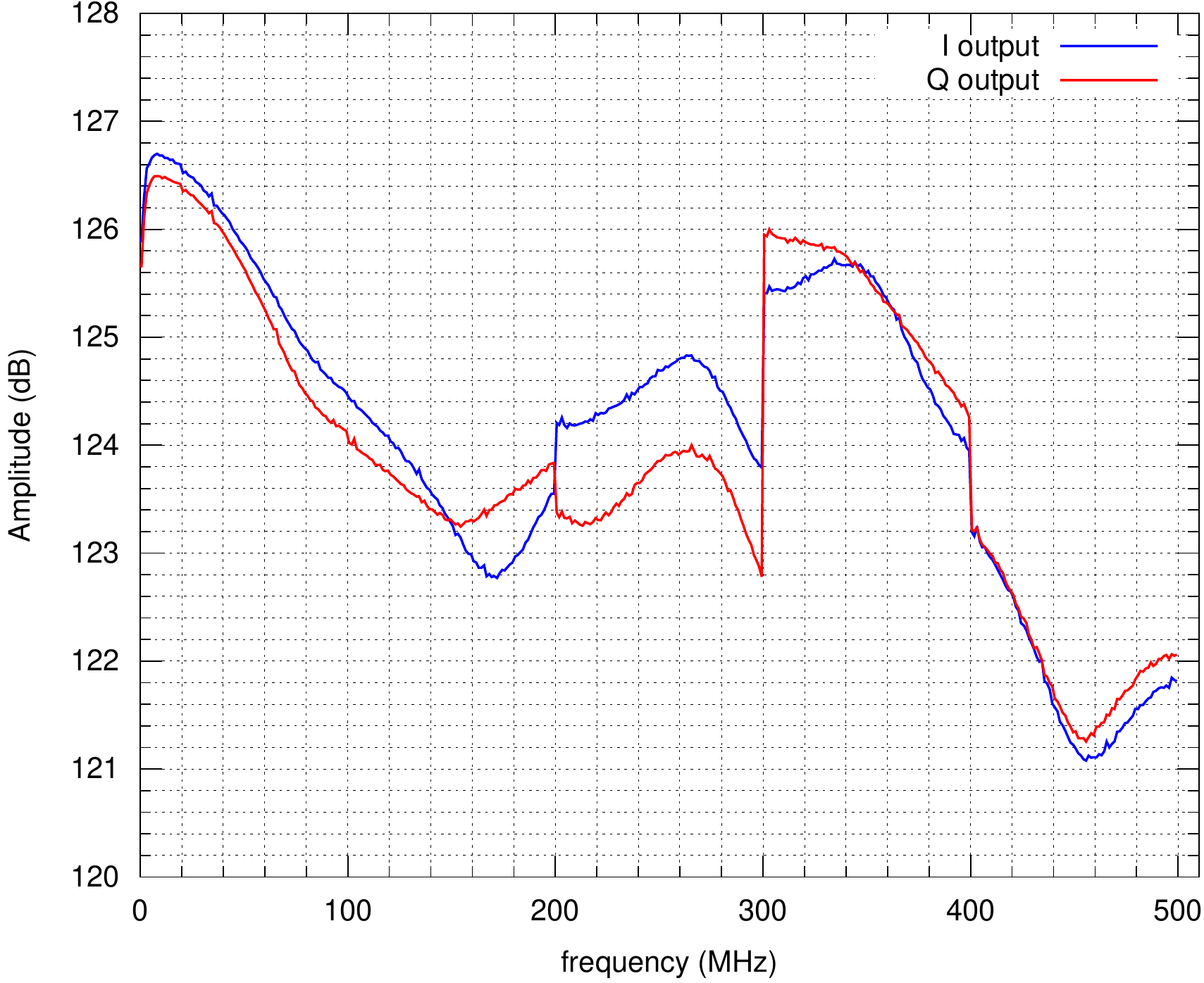}
\caption{Plot of the system frequency response measured for In phase and Quadrature output of the board in loop-back mode, i.e one of the board output connected directly to the board ADC input. For each measurement, 400 tones uniformly distributed over the system bandwidth were generated and analyzed by the embedded DDC.}
\label{freqResponse}
\end{center}
\end{figure}

The frequency response of the system was measured for In phase and Quadrature output of the board in loop-back mode, i.e one of the board output connected directly to the board ADC input. For each measurement, 400 tones uniformly distributed over the system bandwidth were generated and analyzed by the embedded DDC. The amplitude of each tone is plotted in figure~\ref{freqResponse}.

The expected juxtaposition of the five frequency bands of 100\,MHz, corresponding to each DAC contribution, can be observed on the plot. The maximum amplitude variation observed over the full bandwidth is less than 6\,dB.

We explain the amplitude variation by several factors. A part of the dispersion is due to the active and passive electronic components that display a certain amount of dispersion. For instance, the DAC gain has a worst case dispersion  of $\pm$3.6\,\%, while the DAC full scale current resistor has a dispersion of $\pm$1\,\%. Then, there are also the dispersion of the resistor in the passive combiner and the balun transformer loss dispersion (not documented). Additionally, the balun transformers have a frequency dependent loss (-2\,dB at 500\,MHz) which partly explains the decreasing tendency of the curve. It may also be noted, that the original board design was foreseen to use sum amplifiers to sum the five DAC signals (I and Q). Unfortunately they were causing distortion and picking noise from the power supplies. In consequence, they  were replaced by passive combiners. This modification required the implementation of wire straps to bypass the amplifiers that certainly induces attenuation as the frequency 
increase.

Besides, the DAC output of the bands 100-300\,MHz and 400-500\,MHz were not routed on the outer PCB layers (as striplines), but in the inner FR4 layers (as microstrips) and thus they have higher dielectric losses. From the dielectric manufacturer specification, a loss difference of 0.2\,dB can be observed between FR4 and RO4350 microstrips.

Finally, some routing choices were not optimal (bends, stubs, ...) and certainly, they cause small impedance variations over the lines which induce transmission losses as well.

Even though the fluctuation is not fully explained, it remains totally acceptable for such a bandwidth. Moreover, this can be corrected by applying tone per tone power adjustment.

\subsection{System noise}
As shown in \cite{Bourrion2011}, the main system noise contributors in the KID readout electronic chain are the RF mixing electronics and the cold amplifier.
Consequently, this prototype was also tested in loop-back to measure its noise power spectrum distribution. The measurements were performed for one tone generated in the middle of each  frequency band and at different output power level. 
The output level was digitally adjusted with the digital gain module available in each FPGA `proc' (see figure~\ref{FPGA_proc_nikel}).
The highest signal level reached by this method was just slightly above midscale for the 2\textsuperscript{5} gain.

For each tone and in each digital gain conditions, 6000 points were recorded at 23.84\,Hz and were windowed with a Hann function. 
The resulting data were used for computing the Fast Fourier Transform (FFT) and the 6\,dB loss due to the windowing function was compensated. Finally the resulting FFT was smoothed by FFT filtering (20 bins kept).

\begin{figure}[th]
\begin{center}
\includegraphics[angle=0,width=0.8\textwidth]{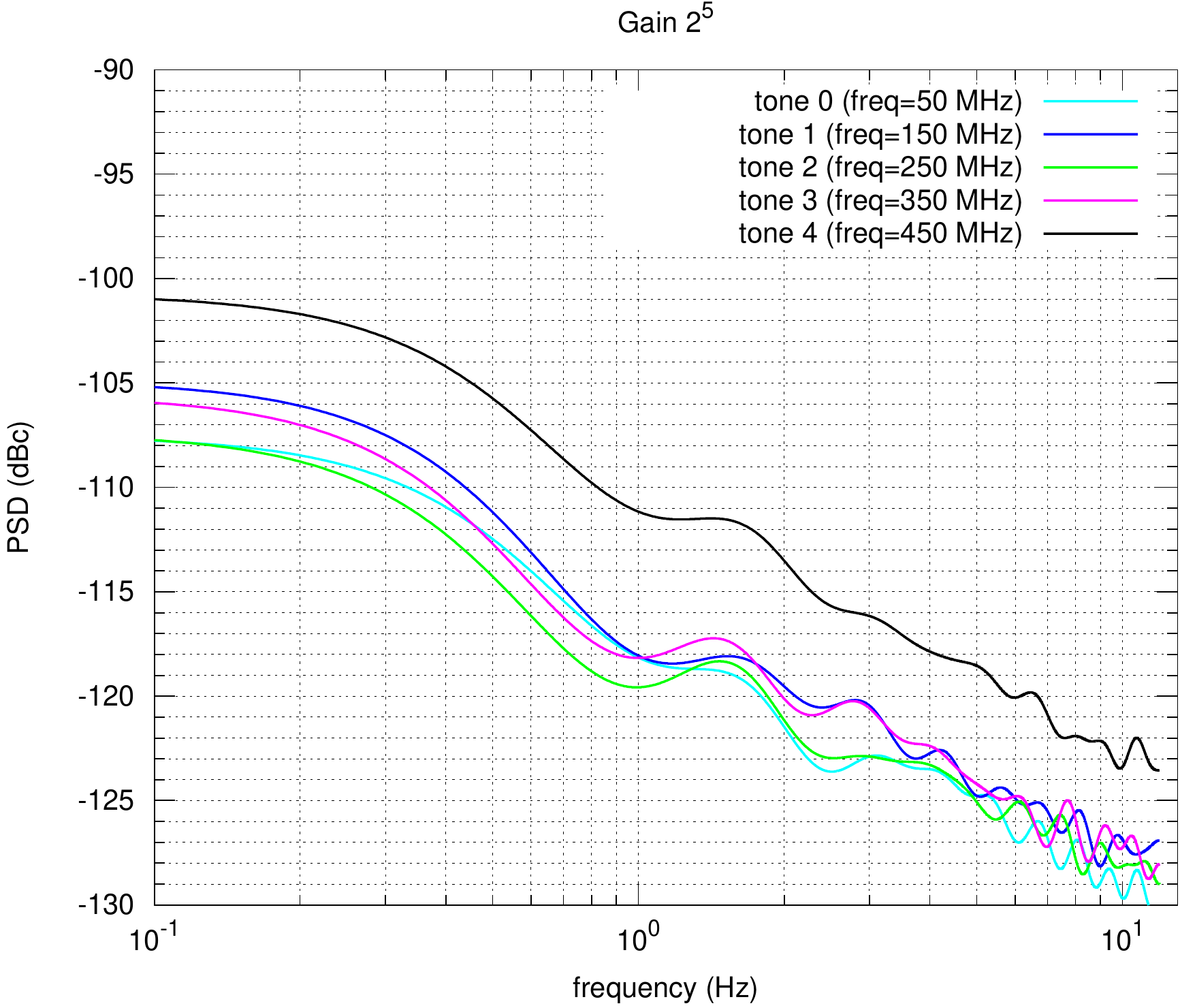}
\caption{Power spectrum distribution plot showing system noise for one tone in each band. At the exception of tone 4, all tone have a similar Signal to Noise Ratio (SNR).}
\label{allTone_Gain5}
\end{center}
\end{figure}

Figure~\ref{allTone_Gain5} shows the system noise Power Spectrum Distribution (PSD) for one tone in each frequency band. 
With the exception of tone 4, all tones have a similar Signal to Noise Ratio (SNR). This is compatible with the board losses mentioned previously that reduce the signal amplitude by about 6\,dB.

\begin{figure}[th]
\begin{center}
\includegraphics[angle=0,width=0.8\textwidth]{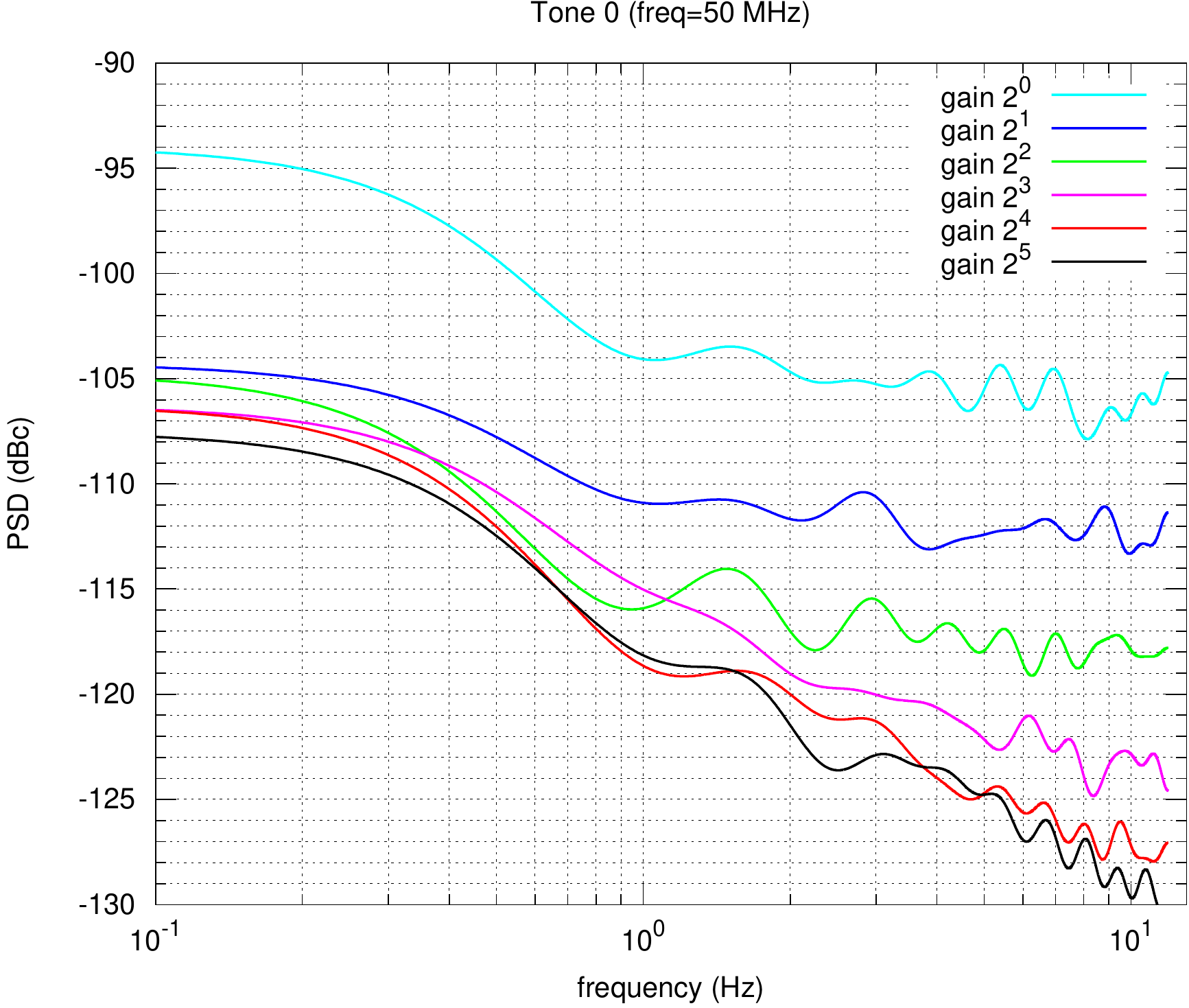}
\caption{Power spectrum distribution for a given tone but for different signal amplitude. The noise floor (relative to carrier) can be seen to be increased accordingly to each amplitude decrease.}
\label{allGain_Tone0}
\end{center}
\end{figure}

Figure~\ref{allGain_Tone0} shows the system noise PSD for a given tone but for different excitation signal amplitudes. The noise floor (relative to carrier) is seen to increase accordingly with each amplitude decrease. 
It may be noticed that when all tones are activated in a single band, it is possible to keep a digital gain between 2\textsuperscript{1} and  2\textsuperscript{2} without DAC clipping because of the frequency values random distribution which minimizes the risk to sum all tones at their maximum amplitude at the same time. Therefore, the 2\textsuperscript{1} and  2\textsuperscript{2} gain curves, provide the achievable performance when the full capabilities of the board are used.

\section{Conclusion}
We have presented in this paper a first prototype of the NIKEL electronic board which was specifically designed
for the NIKA camera to be installed at the IRAM 30\,m telescope at Pico Veleta, Spain. 
We have proved that NIKEL is able to perform real-time frequency multiplexing of an array of up to 400 MKIDs over a bandwidth of 500\,MHz
with outstanding performances in terms of noise. This is due to an innovative solution based on the splitting of the original 500\,MHz band into
five bands of 100\,MHz each, thanks to state of art electronic components and sophisticated numerical filtering algorithms.
The NIKEL multiplexing factor is three times larger compared to previous single board systems and it opens a clear path towards the exploitation and monitoring of future kilo-pixel arrays of MKIDs. Consequently, the resulting minimization of the cable count towards the cryogenic system makes it an asset. Such large arrays will be with no doubt a serious alternative to standard bolometric techniques for millimeter astronomy both because of the intrinsic quality of MKIDs (low noise and fast response) and because of the large multiplexing capabilities.


\end{document}